# Option pricing using a skew random walk pricing tree

Yuan Hu[1], W. Brent Lindquist[2,*], Svetlozar T. Rachev[2], Frank J. Fabozzi[3]

[1] Department of Mathematics, University of California San Diego, La Jolla, CA USA

[2] Department of Mathematics & Statistics, Texas Tech University, Lubbock, TX USA

[3] Finance Group, EDHEC Business School, Nice, Fr.

[*] Corresponding author: brent.lindquist@ttu.edu

**Abstract.** Motivated by the Corns-Satchell, continuous time, option pricing model, we develop a binary tree pricing model with underlying asset price dynamics following Itô-Mckean skew Brownian motion. While the Corns-Satchell market model is incomplete, our discrete time market model is defined in the natural world; extended to the risk neutral world under the no-arbitrage condition where derivatives are priced under uniquely determined risk-neutral probabilities; and is complete. The skewness introduced in the natural world is preserved in the risk neutral world. Furthermore, we show that the model preserves skewness under the continuous-time limit. We provide numerical applications of our model to the valuation of European put and call options on exchange-traded funds tracking the S&P Global 1200 index.

# 1. Introduction

In the academic literature on dynamic asset pricing theory, discrete-time option pricing models[1] traditionally play a subordinated role to their continuous-time counterparts[2]. This is due to the highly developed theory of semimartingales[3], Lévy processes[4] and the fundamental theorem of asset pricing[5]. However, as discussed in Section 2, the standard option pricing approach, based upon passing to the continuous-time limit results in the loss of significant information about the stylized-fact capturing parameters embedded in the discrete-time market models and a consequent dislocation between the information on spot prices that is available to spot traders versus option traders.[6] We further note that pure continuum models that attempt to incorporate stylized facts such as the heavy-tails and skewness properties of realistic price processes through the use of semimartingale (e.g. Lévy) processes, generally lead to incomplete market models. The choice of the equivalent risk-neutral martingale becomes ad hoc, and any unique relationship between the probability measures driving price processes in the natural and risk-neutral worlds is lost.

---

[1] See: Brennan and Schwartz (1978); Rendleman and Bartter (1979); Cox, Ross and Rubinstein (1979); Jarrow and Rudd (1983, Chapters V & VI); Omberg (1987, 1988), Boyle (1988); Boyle, Evnine and Gibbs (1989); He (1990); Nelson and Ramaswamy (1990); Amin (1991); Pedersen and Shiu (1993); Boyle and Lau (1994); Rogers and Stapleton (1997); Rubinstein (1998); Tian (1999); Chang and Palmer (2007); Korn and Korn (2000, Chapter 4); Walsh (2003); Korn and Müller (2010); Simonato (2011); Leccadito, Toscano and Tunaru (2012); Yuen, Siu and Yang (2013); and Bock and Korn (2016).

[2] See: Black and Scholes (1973); Merton (1973, 1976); Cox and Ross (1976); Harrison and Kreps (1979); Harrison and Pliska (1981, 1983); Bensoussan (1984); Hull and White (1987); Karatzas (1988); Karatzas and Shreve (1991); Madan and Milne (1991); Heston (1993); Karatzas and Kou (1996); Louis (1997); Cambell, Lo and MacKinlay (1997, Chapter 9); Madan, Carr and Chang (1998); Chan (1999); Fouque, Papanicolaou and Sircar (2000); Duffie (2001); Kou (2002); Markus and Wu (2002); Pan (2002); Schoutens (2003); Bjoerk (2004); Hunt and Kennedy (2004); Bingham and Kiesel (2004); Shreve (2004); Carr and Wu (2004); Applebaum (2004); Huang and Wu (2004); Elliott and Kopp (2005); Eberlein, Papapantoleon and Shiryaev (2008); Barndorff-Nielsen and Stelzer (2011); Rachev et al. (2011); Miyahara (2012); Yeap, Kwok and Choy (2018); and Shirvani et al (2020).

[3] See Émery and Yor (2002); Jacod and Shiryaev (2003); and Protter (2004).

[4] See Bertoin (1996); Sato (1999); Barndorff-Nielsen, Mikosch and Resnick (2001); Applebaum (2004).

[5] See Delbaen and Schachermayer (1994, 1998); Elliott and Kopp (1999, Chapter 3); Guasoni, Lépinette and Rásonyi (2012); and Jarrow (2012).

[6] In our discrete option pricing model, there is a unique one-to-one mapping between the risk-neutral and natural probability measures. Thus, option traders can calibrate the risk-neutral measure from option data and pass to the unique natural measure to see whether the "implied spot prices" significantly differ from the market spot prices. In particular, option traders can calibrate the "implied" mean instantaneous return and compare it with the instantaneous return derived from the spot price data (Ross, 2013; Audrino, Huitema and Ludwig, 2014). These techniques for determining market dislocation, i.e., the discrepancies in the views of option traders on spot prices and the market values of spot prices, were used during the market crash of 2008. However, in incomplete markets where there is no unique one-to-one mapping of the risk-neutral and natural measures, determining such market dislocation is not possible.

This motivates us to propose a primary role for discrete-time option pricing models, and that any continuous-time option pricing model should only be considered[7] as a benchmark model for a more precise discrete pricing model that preserves valuable market characteristics. Based on this proposal, we identify three fundamental research issues. The first is to develop pricing trees that capture valuable market characteristics in the natural world and retain these in the risk-neutral world. The second is to ensure that the model is market complete. In recognition of the role that continuous-time benchmark models play, the third would be to determine whether the pricing tree retains these characteristics in the limit that the transaction time interval $\Delta t \downarrow 0$.

These research issues have been the focus of a series of papers (Hu et al, 2020a, 2020b, 2021). In Hu et al. (2021), the authors developed a market complete option valuation based on a generalization of the Jarrow-Rudd (JR) pricing model (1983). In contrast to other similar works, all of which placed the binomial tree model directly in the risk-neutral world (and were therefore unable to state what the determining tree is in the natural world), Hu et al. begin in the natural world and, from that, derive the unique (and hence market-complete) risk-neutral tree. Their generalization involved the use of skew Brownian motion (SBM) (Itô-McKean, 1996) with the incorporation of a parameter $\beta$ (in addition to the mean $\mu$ and standard deviation $\sigma$) to capture skewness and kurtosis effects in the real-world pricing tree. These three parameters, $\mu, \sigma$ and $\beta$, carry over into the risk-neutral world. Using the Cherny-Shirayaev-Yor invariance principle (2003), Hu et al. (2021) showed that for a fixed, but relatively small, time increment, $\Delta t$, the pricing tree behavior was geometric SBM (GSBM). However, in the limit $\Delta t \downarrow 0$, the model converged to geometric BM (GBM), losing the skewness and kurtosis properties of the discrete model.

---

[7] The implied volatility, a parameter used to fit continuous-time option pricing models to existing market data, is a case in point. It fully deserves the aphorism: "The implied volatility is the wrong number that, used in the wrong formula, gives the right price for the option." (Marroni and Perdomo, 2014, p. 123; Joshi (2008, p. 197), and Rebonato (2004, p. 169)).

Corns and Satchell (2007) developed a continuous-time pricing model combining a standard BM and a SBM to capture skewness in option pricing models. Unfortunately, their model, as well as the model by Zhu and He (2017) which provides corrections to the Corns-Satchell (CS) model, is market incomplete since the equivalent martingale obtained in their work only provides hedging against the risks associated with the standard BM. These two papers serve as motivation for this article on discrete-option pricing.

In this paper, beginning in the natural world with a skewed, heavy-tailed, binomial process, we derive the unique, market complete, risk-neutral pricing tree that preserves these features. We then show that the continuum-time model preserves these properties. The "cost" to achieve this is the addition of a second Brownian motion to the discrete pricing process. As a result, we show that the risk-free rate in the risk-neutral process cannot be arbitrary, but is a rate determined by an appropriate combination of the means and variances governing the returns of the underlying risky assets.

More specifically, we introduce a binary pricing tree, which we refer to as the Itô-McKean skew pricing tree (IMSPT), that can model both standard BM and Itô-McKean SBM. As we ensure hedging against the risks associated with both the standard BM and the SBM in the model, the IMSPT will lead to a complete market model for option pricing. In the IMSPT, the mean return of the underlying stock is preserved in the resulting option pricing model[8]. The role of the uptick probability $p_{\Delta t}$ governing the drift of a standard binomial price process (see Section 2) is replaced by a skewness parameter $\alpha \in (0,1)$ in the IMSPT. The skewness parameter $\alpha$ remains present in both the natural and the risk neutral world. We show continuous time convergence of the IMSPT model to skew GBM, preserving the skew properties of the pricing tree.

---

[8] In the CS option pricing model, as in any continuous time option pricing model, the mean stock return parameter is not present.

The paper is organized as follows. In Section 2 we elucidate the standard four-step approach used in binomial option pricing that results in the loss of critical natural world information. In Section 3 we list properties of Itô-McKean SBM, while Section 4 introduces our complete market pricing tree model. We apply the Harrison-Shepp functional limit theorem to show the weak convergence of our pricing tree to GSBM. We provide a numerical example that fits the resulting GSBM to return values based upon daily closing prices of the SPDR S&P 500 ETF Trust. In Section 5 we construct an option pricing model based on the IMSPT for the minimum number of underlying risky assets needed to ensure a complete market model. We provide a numerical example to price European call and put options having underlying exchange-traded funds (ETFs) that track component indices of the S&P Global 1200 Index.

## 2. The standard approach − loss of information in passing to the continuous time limit

Consider the standard four-step approach to binomial option pricing[9].

**Step 1.** A binomial pricing tree $\{S_{k\Delta t}, \ k = 0, \ldots, n, \ n\Delta t = T\}$ is introduced in the natural world $\mathbb{P}$. The price dynamics of the underlying risky asset (stock) follows a binomial tree

$$S_{(k+1)\Delta t} = \begin{cases} S_{k\Delta t} u_{\Delta t} & \text{w. p.} \quad p_{\Delta t}, \\ S_{k\Delta t} d_{\Delta t} & \text{w. p.} \quad 1 - p_{\Delta t}, \end{cases}, \qquad p_{\Delta t} \in (0,1), \quad k = 0, \ldots, n-1$$

conditionally on $S_{k\Delta t}$. The no-arbitrage condition requires that $u_{\Delta t} > e^{r\Delta t} > d_{\Delta t}$, where $r \geq 0$ is the instantaneous riskless rate. The log-return $R_{(k+1)\Delta t} = \ln(S_{(k+1)\Delta t}/S_{k\Delta t})$ is assumed to have a mean $\mathbb{E}(R_{(k+1)\Delta t}) = \mu \Delta t$ and a variance $\text{Var}(R_{(k+1)\Delta t}) = \sigma^2 \Delta t$ to guarantee that the binomial pricing tree has a weak Gaussian limit (needed in Step 2). With $o(\Delta t) = 0$, the pair $(u_{\Delta t}, d_{\Delta t})$ is then uniquely determined[10] as $u_{\Delta t} = 1 + \mu\Delta t + \sqrt{(1-p_{\Delta t})/p_{\Delta t}}\,\sigma\sqrt{\Delta t}$ and $d_{\Delta t} = 1 + \mu\Delta t - \sqrt{p_{\Delta t}/(1-p_{\Delta t})}\,\sigma\sqrt{\Delta t}$.

---

[9] See, for example: Cox and Rubinstein (1985, Chapter 5); Korn and Korn (2000, Section 4.3); Bingham and Kiesel (2004, Chapter 4); Hull (2012, Chapters 12 & 20); and Marroni and Perdomo (2014, Chapter 4).

**Step 2**. From Step 1 and the Donsker-Prokhorov invariance principle (DPIP)[11], the price process generated by the tree in Step 1 converges weakly in the Skorokhod $\mathfrak{D}[0,T]$-space to a GBM, $S_t = S_0 \exp\big((\mu - \sigma^2/2)t + \sigma B_t\big)$, $t \in [0,T]$, where $B_t$ is a BM on $\mathbb{P}$. With this step, the information about the probability $p_{\Delta t}$ for stock-price uptick that was available in the discrete-time process has been lost.

**Step 3.** Using the no-arbitrage condition on the continuum model from Step 2, the dynamics of the underlying asset in the risk-neutral world $\mathbb{Q}$ is $S_t = S_0 \exp\big((r - \sigma^2/2)t + \sigma B_t^{\mathbb{Q}}\big)$, $t \in [0,T]$, where $B_t^{\mathbb{Q}}$ is a BM on $\mathbb{Q}$. Now a hedger, who takes the short position in an option contract and who, in any real trading situation, will trade only in discrete time instances, has lost information about both the instantaneous mean return $\mu$ and $p_{\Delta t}$. The loss of $\mu$ is due to the absurd assumption that the trader can trade continuously in time. Including trading costs in continuous-time option pricing[12] cannot retrieve the information lost about $p_{\Delta t}$ and $\mu$. No real trader can trade in continuous time, with or without transaction costs.

**Step 4.** Step 3 and the DPIP imply that the risk-neutral, continuous asset price dynamics can be approximated by a risk-neutral binomial tree with price dynamics given by

$$S_{(k+1)\Delta t} = \begin{cases} S_{k\Delta t} u_{\Delta t}^{(\mathbb{Q})} & \text{w.p.} \quad p_{\Delta t}^{(\mathbb{Q})} \\ S_{k\Delta t} d_{\Delta t}^{(\mathbb{Q})} & \text{w.p.} \quad 1 - p_{\Delta t}^{(\mathbb{Q})} \end{cases}, \quad k = 0, \dots, n-1,$$

---

[10] See Rendleman and Bartter (1979).
[11] See: Donsker (1951); Prokhorov (1956); Billingsley (1999, Section 14); Gikhman and Skorokhod (1969, Chapter IX); Skorokhod (2005, Section 5.3.3); and Davydov and Rotar (2008).
[12] See Leland (1985); Hodges and Neuberger (1989); Boyle and Vorst (1992); Davis, Vassilios and Zariphopoulou (1993); Edirsinghe, Naik and Uppal (1993); Broadie, Cvitanic and Soner (1998); Kabanov and Stricker (2001); Zakamouline (2009); Lai and Lim (2009); and Guasoni, Lépinette and Rásonyi (2012).

where

$$u_{\Delta t}^{(\mathbb{Q})} = 1 + r\Delta t + \sqrt{\left(1 - p_{\Delta t}^{(\mathbb{Q})}\right)/p_{\Delta t}^{(\mathbb{Q})}}\, \sigma\sqrt{\Delta t} \quad \text{and} \quad d_{\Delta t}^{(\mathbb{Q})} = 1 + r\Delta t - \sqrt{p_{\Delta t}^{(\mathbb{Q})}/\left(1 - p_{\Delta t}^{(\mathbb{Q})}\right)}\, \sigma\sqrt{\Delta t}$$

for any arbitrarily chosen value of $p_{\Delta t}^{(\mathbb{Q})} \in (0,1)$. Setting

$$p_{\Delta t}^{(\mathbb{Q})} = p_{\Delta t}^{(\mathbb{Q};CRR)} \equiv \frac{e^{r\Delta t} - e^{-\sigma\sqrt{\Delta t}}}{e^{\sigma\sqrt{\Delta t}} - e^{-\sigma\sqrt{\Delta t}}} = \frac{1}{2} + \frac{r - \sigma^2/2}{2\sigma}\sqrt{\Delta t}$$

results in the Cox-Ross-Rubinstein (CRR) pricing tree[13]. Setting $p_{\Delta t}^{(\mathbb{Q})} = p_{\Delta t}^{(\mathbb{Q};JR)} = 1/2$ results in the JR pricing tree[14]. By the DPIP, the value of $p_{\Delta t}^{(\mathbb{Q})} \in (0,1)$ controls the rate of convergence of the binomial option price to its continuous time limit[15] and is otherwise irrelevant in that limit. Thus, returning to the realistic case when the hedger trades in discrete time instances, the pricing tree (which the hedger is supposed to use to hedge his short position in the option contract) is now without the parameters $\mu$ and $p_{\Delta t}$. This, in turn, leads to a dislocation in the information available to spot traders and to option traders with regards to spot prices. In real option trading, with hedging done in discrete time instances, $k\Delta t$, $k = 0, 1, \ldots, n-1$, no option trader will disregard information about $\mu$ and $p_{\Delta t}$.

---

[13] See Cox, Ross and Rubinstein (1979). If $p_{\Delta t}^{(\mathbb{Q})} = p_{\Delta t}^{(\mathbb{Q};CRR)}$, then since $p_{\Delta t}^{(\mathbb{Q};CRR)} = p_{\Delta t} - \theta\sqrt{p_{\Delta t}(1 - p_{\Delta t})}\sqrt{\Delta t}$ (and under the condition $o(\Delta t) = 0$), the probability for stock price upturn in the natural world is uniquely determined as $p_{\Delta t} \equiv p_{\Delta t}^{(\mathbb{P},CRR)} = \frac{1}{2} + \frac{|\mu - \sigma^2/2|}{2\sigma}\sqrt{\Delta t}$. The value of $p_{\Delta t}$ that is historically observed is generally different from $p_{\Delta t}^{(\mathbb{P},CRR)}$.

[14] See Jarrow and Rudd (1983, Chapters V & VI) and Hull (2012, p. 442). If $p_{\Delta t}^{(\mathbb{Q})} = p_{\Delta t}^{(\mathbb{Q};JR)}$, then since $p_{\Delta t}^{(\mathbb{Q};JR)} = p_{\Delta t} - \theta\sqrt{p_{\Delta t}(1 - p_{\Delta t})}\sqrt{\Delta t}$ (and under the condition $o(\Delta t) = 0$), the probability for stock price upturn in the natural world is uniquely determined as $p_{\Delta t} \equiv p_{\Delta t}^{(\mathbb{P};JR)} = \frac{1}{2} + \frac{1}{2}|\theta|\sqrt{\Delta t}$. The value of $p_{\Delta t}$ that is historically observed is generally different from $p_{\Delta t}^{(\mathbb{P};JR)}$.

[15] When $p_{\Delta t}^{(\mathbb{Q})}$ approaches 1 or 0, the rate of convergence becomes slower and vanishes when $p_{\Delta t}^{(\mathbb{Q})} \uparrow 1$ or $p_{\Delta t}^{(\mathbb{Q})} \downarrow 0$. As shown in Kim et al. (2016), the rate of convergence (in a suitably chosen metric) is fastest when $p_{\Delta t}^{(\mathbb{Q})} = 1/2$.

**Remark 1:** There is no need to pass to continuous time price dynamics when using binomial option pricing[16,17]. Proceeding immediately from Step 1 to the discrete risk-neutral pricing tree, the no-arbitrage argument implies that the risk-neutral probability, $q_{\Delta t}$, over the interval $[k\Delta t, (k+1)\Delta t]$, $k = 0, \ldots, n-1$, is uniquely determined by $q_{\Delta t} = p_{\Delta t} - \theta\sqrt{p_{\Delta t}(1-p_{\Delta t})}\sqrt{\Delta t}$, where $\theta = (\mu - r)/\sigma$ is the market price of risk[18]. The risk-neutral asset pricing tree is

$$S_{(k+1)\Delta t} = \begin{cases} S_{k\Delta t} u_{\Delta t} & w.p. \quad q_{\Delta t}, \\ S_{k\Delta t} d_{\Delta t} & w.p. \quad 1 - q_{\Delta t}, \end{cases} \quad k = 0, \ldots, n-1.$$

The four parameters $\mu, \sigma, r$ and $p_{\Delta t}$ are preserved in the risk-neutral price process.[19] The resultant binomial market model is an arbitrage-free, complete market model. Therefore, if the option trader had bypassed Steps 2, 3 and 4 and passed directly from Step 1 to the discrete risk-neutral pricing tree, information about $\mu$ and $p_{\Delta t}$ would be preserved when valuing the option price with the uniquely determined risk neutral probabilities given by $q_{\Delta t} = p_{\Delta t} - \theta\sqrt{p_{\Delta t}(1-p_{\Delta t})}\sqrt{\Delta t}$ and $1 - q_{\Delta t}$.

**Remark 2:** Trinomial and multinomial pricing trees appearing in the literature are placed directly in the risk neutral world, bypassing Steps 1, 2 and 3 altogether[20]. These option pricing trees in complete market models do not address the critical question as to which uniquely determined pricing tree in the natural world leads to the selected trinomial or multinomial pricing tree in the risk neutral world.

---

[16] For various generalizations of the binomial pricing market model in Step 1, see Hu et al. 2020a, 2020b, and 2021.

[17] See also Jarrow, Protter and Sayit (2009). The idea is to disallow continuous trading and, moreover, to require a minimal fixed time between successive trades. This fixed time can be as small as one likes, but once chosen, it cannot be changed. This disallows a clustering of trades around a fleeting arbitrage opportunity, such as might occur from a drift process that the random generating process cannot "see."

[18] See Kim et al (2016, 2019).

[19] The trader's information about $p_{\Delta t}$ and $\mu$, could be gathered from stock market data or be based on trader's active trading alpha-strategy; see, for example, Grinold and Kahn (2000, Chapter 4).

[20] See Boyle (1986, (1988), Boyle, Evnine, and Gibbs (1989), Madan, Milne, and Shefrin (1989), Kamrad and Ritchen (1991), Boyle and Lau (1994), Florescu and Viens (2008), Deutsch (2009, Chapter 10), and Ma and Zhu (2015).

## 3. Itô-McKean skew Brownian motion

A SBM[21] $\mathbb{A}^{(\delta)}$ is defined as follows. Let $\mathbb{B} = \{B_t, t \geq 0\}$ be a standard BM generating a stochastic basis $(\Omega, \mathbb{F} = (\mathcal{F}_t = \sigma(B_u, u \leq t); t \geq 0), \mathbb{P})$. Let $\alpha \in (0,1)$ and set

$$\mathcal{A}_t^{(\alpha)} = \int_0^t (\alpha^2 I_{\{B_s \geq 0\}} + (1-\alpha)^2 I_{\{B_s < 0\}}) \, ds,$$

$$\tau_t^{(\alpha)} = \inf\{s \geq 0 : \mathcal{A}_s^{(\alpha)} > t\}, \qquad t \geq 0,$$

$$B_t^{(\alpha)} = \varphi_\alpha\left(B_{\tau_t^{(\alpha)}}\right), \qquad t > 0, \qquad B_0^{(\alpha)} = 0,$$

where $\varphi_\alpha(x) = \alpha x I_{\{x \geq 0\}} + (1-\alpha) x I_{\{x < 0\}}$, $x \in R$. The process $\mathbb{B}^{(\alpha)} = \{B_t^{(\alpha)}, t \geq 0\}$ is called a *SBM with parameter $\alpha$*.[22] The SBM $\mathbb{B}^{(\alpha)}$ is a semimartingale satisfying the strong Markov property[23] and $\mathbb{B}^{(\frac{1}{2})} \stackrel{d}{=} \mathbb{B}$. For every $t \geq 0$, the density $f_t^{(\alpha)}(x)$, $x \in R$ of $B_t^{(\alpha)}$ is given by

$$f_t^{(\alpha)}(x) = \begin{cases} \alpha \sqrt{\dfrac{2}{\pi t}} \exp\left(-\dfrac{x^2}{2t}\right), & \text{if } x \geq 0, \\ (1-\alpha) \sqrt{\dfrac{2}{\pi t}} \exp\left(-\dfrac{x^2}{2t}\right), & \text{if } x < 0. \end{cases}$$

As noted in Cherny, Shirayaev, and Yor (2003), there are other ways of defining a SBM (all of which are equal in distribution). Trajectories of $\{B_t^{(\alpha)}, t \geq 0\}$ can be simulated using the representation

---

[21] See §4.2, Problem 1, p. 115, in Itô and McKean (1996, Reprint of the 1974 edition). See also: Harrison and Shepp (1981); Revuz and Yor (1994, Chapters VII and X); Lang (1995); Lejay (2006); Corns and Satchell (2007); Cherny, Shiryaev, and Yor (2003); Ramirez (2011); Atar and Budhiraja (2015); Trutnau, Ouknine. and Russo (2015); and Li (2019).

[22] We follow the definition of a skew Brownian motion in Cherny, Shiryaev, and Yor (2003).

[23] See Cherny, Shiryaev, and Yor (2003, Section 4) and Corns and Satchell (2007).

$$B_t^{(\alpha,\text{IM})} = \begin{cases} |B_t| & \text{w. p. } \alpha, \\ -|B_t| & \text{w. p. } 1-\alpha, \end{cases}$$

where, as noted above, $B_t$ is a standard BM. The above representation first appears in Itô and McKean (1996, §4.2, Problem 1, page 115). Corns and Satchell (2007) (CS) developed a representation which they called Itô-Mckean SBM. The Itô-Mckean SBM $\mathbb{A}^{(\delta)} = \left\{A_t^{(\delta)}, t \geq 0\right\}$ with parameter $\delta \in (-1,1)$ is defined as

$$A_t^{(\delta)} = \sqrt{1-\delta^2}B_{1,t} + \delta|B_{2,t}|, \tag{1}$$

where $\mathbb{B}_1 = \{B_{1,t}, t \geq 0\}$ and $\mathbb{B}_2 = \{B_{2,t}, t \geq 0\}$ are two independent BMs. CS showed that $\mathbb{A}^{(\delta)}$ has the conditional probability density

$$\mathbb{P}\left(A_t^{(\delta)} \in dz\right) = \frac{2}{\sqrt{t}}\varphi\left(\frac{A_t^{(\delta)}}{\sqrt{t}}\right)\Phi\left(\lambda\frac{A_t^{(\delta)}}{\sqrt{t}}\right)dz,$$

where $\lambda = \delta/\sqrt{1-\delta^2}$, and $\varphi(\cdot)$ and $\Phi(\cdot)$ are, respectively, the standard normal density and cumulative distribution functions. CS showed that $\mathbb{A}^{(\delta)} \stackrel{d}{=} \mathbb{B}^{(\alpha,\text{IM})}$ with $\alpha = (1+\delta)/2$. The density $f_{Z^{(\lambda)}}(z) = 2\phi(z)\Phi(\lambda z)$ governing a continuous random variable $Z^{(\lambda)}$ is referred to as a skewed normal distribution (SND). Relevant properties of an SND are presented in Appendix A.

Using the moment properties of an SND, the moments $\mu_t^{(p,\alpha)} = \mathbb{E}\left[\left(B_t^{(\alpha)}\right)^p\right]$, $p > 0$, $t > 0$ are given by $\mu_t^{(p,\alpha)} = \sqrt{2^p/\pi}\,\Gamma((p+1)/2)[\alpha + (-1)^p(1-\alpha)]t^{p/2}$. In particular, the mean, variance, skewness, and excess kurtosis of $B_t^{(\alpha)}$ are given by:

$$\mu_t^{(\alpha)} = \mathbb{E}\left[B_t^{(\alpha)}\right] = (2\alpha - 1)\sqrt{2t/\pi} = \sqrt{t}\mu_1^{(\alpha)};$$

$$\text{Var}\left[B_t^{(\alpha)}\right] \equiv V_t^{(\alpha)} = t\left[1 - \left(\mu_1^{(\alpha)}\right)^2\right] \equiv tV_1^{(\alpha)};$$

$$\gamma_t^{(\alpha)} = \frac{\mathbb{E}\left[\left(B_t^{(\alpha)} - \mu_t^{(\alpha)}\right)^3\right]}{\left(V_t^{(\alpha)}\right)^{3/2}} = \frac{\left[2\left(\mu_1^{(\alpha)}\right)^2 - 1\right]\mu_1^{(\alpha)}}{\left[1 - \left(\mu_1^{(\alpha)}\right)^2\right]^{3/2}} \equiv \gamma_1^{(\alpha)}; \quad (2)$$

$$\kappa_t^{(\alpha)} = \frac{\mathbb{E}\left[\left(B_t^{(\alpha)} - \mu_t^{(\alpha)}\right)^4\right]}{\left(V_t^{(\alpha)}\right)^2} - 3 = \frac{2\left(\mu_1^{(\alpha)}\right)^2\left[1 - 3\left(\mu_1^{(\alpha)}\right)^2\right]}{\left[1 - \left(\mu_1^{(\alpha)}\right)^2\right]^2} \equiv \kappa_1^{(\alpha)}.$$

The terms $\mu_1^{(\alpha)}$, $V_1^{(\alpha)}$, $\gamma_1^{(\alpha)}$ and $\kappa_1^{(\alpha)}$ are introduced in (2) to isolate the time dependence of each moment.

## 4. Itô-McKean SBM pricing tree

Using the Harrison-Shepp functional limit theorem (HSFLT)[24], we elucidate a discrete-time process defined on a finite time interval $[0, T]$ that converges weakly in $\mathcal{D}[0, T]$[25] to Itô-Mckean SBM.[26] Let $\mathbb{M}^{(\alpha)} = \left\{M_k^{(\alpha)} \in \mathbb{Z}, k \in \mathcal{N}_0\right\}$, $\alpha \in (0,1)$, $\mathcal{N}_0 = \{0,1,\ldots\}$, $\mathbb{Z} = \{0, \pm 1, \pm 2,\}$ be a Markov chain, with $M_0^{(\alpha)} = 0$, having the transition probabilities

$$\mathbb{P}\left(M_{k+1}^{(\alpha)} = i + 1 \mid M_k^{(\alpha)} = i\right) = \begin{cases} 1/2 & \text{if } i \neq 0, \\ \alpha & \text{if } i = 0, \end{cases}$$

$$\mathbb{P}\left(M_{k+1}^{(\alpha)} = i - 1 \mid M_k^{(\alpha)} = i\right) = \begin{cases} 1/2 & \text{if } i \neq 0, \\ 1 - \alpha & \text{if } i = 0, \end{cases} \quad (3)$$

for $k \in \mathcal{N}_0$. $\mathbb{M}^{(\alpha)}$ is called a skew random walk (SRW) with parameter $\alpha \in (0,1)$[27]. If $\alpha = 1/2$, $\mathbb{M}^{(\alpha)}$ is a random walk. Relevant properties of the SRW $\mathbb{M}^{(\alpha)}$ are discussed in Appendix B. With $n\Delta t = T$, let $X_{k\Delta t}^{(\alpha)} = \sqrt{\Delta t} M_k^{(\alpha)}$, $k = 1, \ldots, n$, with $X_0^{(\alpha)} = 0$. Consider the following processes in $\mathcal{D}[0,T]$,

$$\mathbb{B}_{[0,T]}^{(\alpha,n)} = \begin{cases} \mathbb{B}_{[0,T]}^{(\alpha,n)}(t) = X_{k\Delta t}^{(\alpha)} & \text{for } t \in [k\Delta t, (k+1)\Delta t), \quad k = 0, \ldots, n-1, \\ \mathbb{B}_{[0,T]}^{(\alpha,n)}(T) = X_{n\Delta t}^{(\alpha)}. \end{cases} \quad (4)$$

---

[24] See Harrison and Shepp (1981) and a more general invariance principle in Cherny, Shiryaev, and Yor (2003).
[25] See Skorokhod (1956) and Billingsley (1999, Chapter 3).
[26] See equation (4). As per the discussion in section 3, all SBMs are equal in distribution. Itô-Mckean SBM is just one representation of an SBM.
[27] See Harrison and Shepp (1981) and Cherny, Shiryaev, and Yor (2003).zS

Then HSFLT states that, as $n \uparrow \infty$, $\mathbb{B}_{[0,T]}^{(\alpha,n)}$ converges weakly in $\mathcal{D}[0,T]$ to the SBM $\mathbb{B}_{[0,T]}^{(\alpha)} = \{B_t^{(\alpha)}, t \in [0,T]\}$.

Consider the pricing tree $S_{k\Delta t}^{(\alpha,n)}, k = 0, 1, \ldots, n$ given by (Hu et al. 2021)

$$S_{k\Delta t}^{(\alpha,n)} = S_0 \exp\left\{\left(\mu_\alpha - \frac{\sigma_\alpha^2}{2}\right) k\Delta t + \sigma_\alpha M_k^{(\alpha)} \sqrt{\Delta t}\right\}, \quad (5)$$

$$\mu_\alpha \in R, \quad \sigma_\alpha \in R, \quad k = 0, \ldots, n, \quad n\Delta t = T,$$

adapted to the filtration $\mathbb{F}^{(\alpha)} = \{\mathcal{F}_k^{(\alpha)} = \sigma(M_j^{(\alpha)}; j = 1, \ldots, k), k \in \mathcal{N}, \mathcal{F}_0^{(\alpha)} = \{\emptyset, \Omega\}\}$. Define the $\mathcal{D}[0,T]$-process

$$\mathbb{S}_{[0,T]}^{(\alpha,n)} = \begin{cases} \mathbb{S}_{[0,T]}^{(\alpha,n)}(t) = S_{k\Delta t}^{(\alpha,n)} & \text{for } t \in [k\Delta t, (k+1)\Delta t), \quad k = 0, \ldots, n-1, \\ \mathbb{S}_{[0,T]}^{(\alpha,n)}(T) = S_{n\Delta t}^{(\alpha,n)}. \end{cases}$$

Then, by HSFLT, $\mathbb{S}_{[0,T]}^{(\alpha,n)}$ converges weakly in $\mathcal{D}[0,T]$ to a GSBM

$$\mathbb{S}_{[0,T]}^{(\alpha)} = \left\{S_t^{(\alpha)} = S_0 \exp\left\{\left(\mu_\alpha - \frac{\sigma_\alpha^2}{2}\right) t + \sigma_\alpha B_t^{(\alpha)}\right\}, \quad t \in [0,T]\right\}. \quad (6)$$

Note that the coefficient $\sigma_\alpha$ appearing in (6) and in its discrete version (5) is not the usual volatility which appears when the driving random motion is Brownian. When the driving motion is Brownian ($\alpha = 1/2$), $\sigma_{1/2} B_t^{(1/2)}$ and $-\sigma_{1/2} B_t^{(1/2)}$ have the same distribution. However, when the driving motion is skew Brownian ($\alpha \neq 1/2$), then $\sigma_\alpha B_t^{(\alpha)}$ and $-\sigma_\alpha B_t^{(\alpha)}$ will not have the same distribution. Thus, when the driving motion is Brownian, the restriction $\sigma_{1/2} > 0$ can be imposed, since any negative sign can be inverted by the change $B_t^{(1/2)} \to -B_t^{(1/2)}$, with no change in the resulting random distribution. For SBM, this is not possible and the sign of $\sigma_\alpha$ is important. Therefore, $\sigma_\alpha$ is not a volatility in the usual sense, but rather a scale parameter, $\sigma_\alpha \in R$, as noted in (5). We refer to $\sigma_\alpha$ as the "marginal scale parameter".

### 4.1 Numerical example: Fitting Itô-Mckean SBM to ETF return values

Fitting the model (3), (5) to data requires identifying best fit values for the parameters $\alpha$, $\mu_\alpha$, $\sigma_\alpha$ over an ensemble of BM trajectories, $M_k^{(\alpha)}$. Appendix C presents a procedure for fitting daily closing prices. The procedure is tested on synthetic data sets produced from (3) and (5) with known values of $\alpha$, $\mu_\alpha$ and $\sigma_\alpha$. The test results show that $\sigma_\alpha$ can be computed very accurately for any single price trajectory. Since the standard deviation of daily returns is larger than the mean value $\mu$, estimation of $\mu$ from a single price trajectory is much less accurate. Estimation of $\alpha$ from a single price trace is slightly better than for estimation of $\mu$; critically, it is possible to infer whether $\alpha$ differs from 0.5, and in what direction.

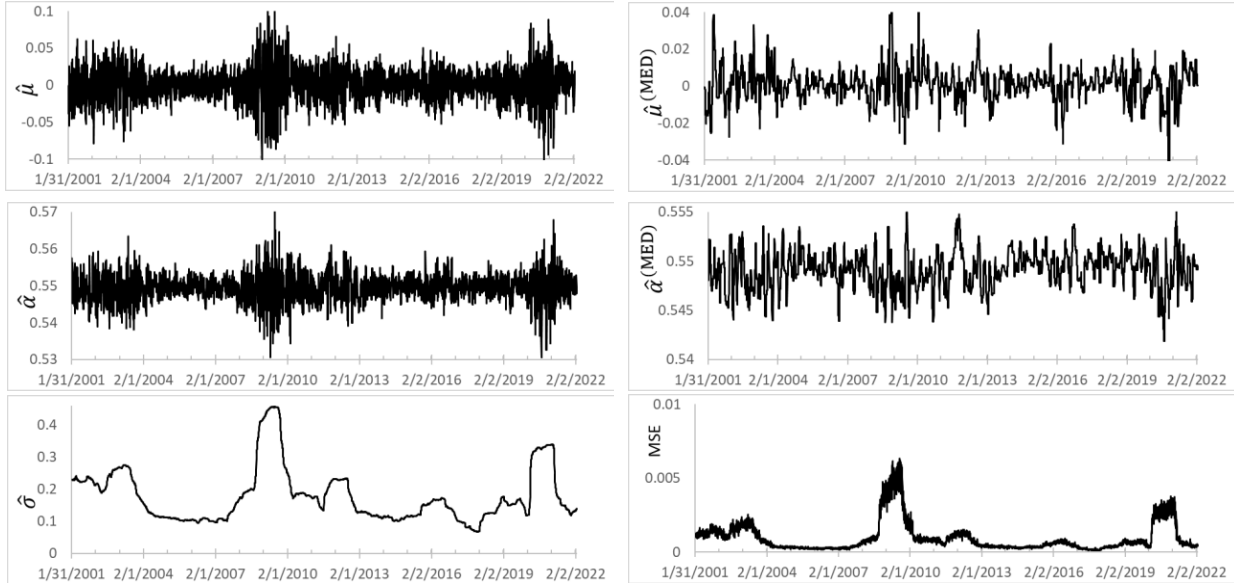

**Figure 1.** Time series of daily optimal values $\hat{\mu}$, $\hat{\alpha}$ and $\hat{\sigma}$ using a 252-day moving window strategy. The time series $\hat{\mu}^{(\text{Med})}$ and $\hat{\alpha}^{(\text{Med})}$ are monthly median values of $\hat{\mu}$ and $\hat{\alpha}$. Also shown is the time series for the mean square error (MSE) for the minimization fit (C4) for $\hat{\mu}$ and $\hat{\alpha}$.

To apply the fitting procedure to a real data set, we consider data[28] for the SPDR S&P 500 ETF Trust (SPY) covering the period 01/03/2000 to 02/17/2022. The results of the fits to the parameters $\hat{\sigma}_t$, $\hat{\alpha}_t$ and $\hat{\mu}_t$ for SPY are given in Fig. 1, using a 252-day moving window. The parameters $\hat{\alpha}_t$ and $\hat{\mu}_t$ were computed using the minimization (C4) with $\sigma^* = \sigma_t^*$ set to the realized average value of $\hat{\sigma}_t$ over each respective window.[29] The time series of mean square errors (MSE) for this minimization are also plotted. To smooth the results for $\hat{\alpha}_t$ and $\hat{\mu}_t$, we computed the median value time series, $\hat{\alpha}_t^{(Med)}$ and $\hat{\mu}_t^{(Med)}$ using a 1-month moving window. The smoothed series are also shown in Fig. 1. The effects of the financial crises of 2008 and the Covid-19 pandemic are discernible in each time series, although they are almost "washed out" of the smoothed time series $\hat{\alpha}_t^{(Med)}$ and $\hat{\mu}_t^{(Med)}$. Mean and standard deviation values for each of these parameter time series are given in Table. 1. To reasonably high precision, $\hat{\alpha}_t$ is consistent with a value of 0.549, indicating that SPY acted as an SBM with positive skewness over this long time period.

**Table 1.** Mean and standard deviation for the time series of the fitted parameters.

|  | $\hat{\sigma}_t$ | $\hat{\mu}_t$ | $\hat{\alpha}_t$ | $\hat{\mu}_t^{(Med)}$ | $\hat{\alpha}_t^{(Med)}$ | MSE |
|---|---|---|---|---|---|---|
| mean | 0.177 | 0.001 | 0.549 | 0.001 | 0.549 | 0.0009 |
| stdev | 0.083 | 0.021 | 0.004 | 0.010 | 0.002 | 0.0010 |

We desire to construct the pricing tree

$$S_{k\Delta t}^{(\hat{\alpha},\hat{\mu},\hat{\sigma},n)} = S_{(k-1)\Delta t}^{(\hat{\alpha},\hat{\mu},\hat{\sigma},n)} \exp\left\{\left(\hat{\mu} - \frac{\hat{\sigma}^2}{2}\right)\Delta t + \hat{\sigma}\sqrt{\Delta t}\left(M_k^{(\hat{\alpha})} - M_{k-1}^{(\hat{\alpha})}\right)\right\}, \qquad (7)$$

---

[28] *Bloomberg Professional Services,* accessed 08/24/2021, 10:00 PM EDT.
[29] As seen from Fig. 1, $\hat{\sigma}_t$ varied by factors of 2 to 5 over this long time period. Using an average value of $\hat{\sigma}_t$ over each moving window for the computation of $\hat{\mu}_t$ and $\hat{\alpha}_t$ produced slightly better results than using a single value of $\sigma^*$, representing an average over the entire time period.

for $k = 1, \ldots, n$, $n\Delta t = T$, with $S_0^{(\hat{\alpha}, \hat{\mu}, \hat{\sigma}, n)} = S_0^{(\text{emp})}$ that approximates an empirical price series $S_{k\Delta t}^{(\text{emp})}$. To do so, we need the "best fit" Markov trajectory $\{M_0^{(\hat{\alpha})} = 0, M_1^{(\hat{\alpha})}, \ldots, M_{k-1}^{(\hat{\alpha})}\}$. Consider the daily log-returns defined by (7),

$$r_{j\Delta t}^{(\hat{\alpha}, \hat{\mu}, \hat{\sigma}, n)} = \ln\left(S_{j\Delta t}^{(\hat{\alpha}, \hat{\mu}, \hat{\sigma}, n)} / S_{(j-1)\Delta t}^{(\hat{\alpha}, \hat{\mu}, \hat{\sigma}, n)}\right) = \left(\hat{\mu} - \frac{\hat{\sigma}^2}{2}\right)\Delta t + \hat{\sigma}\sqrt{\Delta t}\left(M_j^{(\hat{\alpha})} - M_{j-1}^{(\hat{\alpha})}\right),$$

$$j = 1, \ldots, k, \quad r_0^{(\hat{\alpha}, \hat{\mu}, \hat{\sigma}, n)} = 0.$$

This implies that

$$M_j^{(\hat{\alpha})} = M_{j-1}^{(\hat{\alpha})} + \frac{r_{j\Delta t}^{(\hat{\alpha}, \hat{\mu}, \hat{\sigma}, n)} - (\hat{\mu} - \hat{\sigma}^2/2)\Delta t}{\hat{\sigma}\sqrt{\Delta t}}, \quad j = 0, \ldots, k, \quad M_0^{(\hat{\alpha})} = 0. \tag{8}$$

Equating $r_{j\Delta t}^{(\hat{\alpha}, \hat{\mu}, \hat{\sigma}, n)}$ with the observed daily returns, $r_{j\Delta t}^{(\hat{\alpha}, \hat{\mu}, \hat{\sigma}, n)} = r_{j\Delta t}^{(\text{emp})} = R_{j\Delta t}^{(\text{emp})} - R_{j-1\Delta t}^{(\text{emp})}$, $j = 2, 3, \ldots, k$, $r_{1\Delta t}^{(\text{emp})} = R_{1\Delta t}^{(\text{emp})}$, enables computation of a sequence, $M_j^{(\hat{\alpha})}$, $j = 1, \ldots, k$ from (8). However, the values $M_j^{(\hat{\alpha})}$ will (undoubtedly) not be in $\mathbb{Z} = \{0, \pm 1, \pm 2, \ldots\}$. Consequently, we estimate the chain $M_j^{(\hat{\alpha})}$ using the revised formula,

$$\widehat{M}_j^{(\hat{\alpha})} = \widehat{M}_{j-1}^{(\hat{\alpha})} + \text{sign}\left(\frac{r_{j\Delta t}^{(\text{emp})} - (\hat{\mu} - \hat{\sigma}^2/2)\Delta t}{\hat{\sigma}\sqrt{\Delta t}}\right), \quad j = 0, \ldots, k, \quad \widehat{M}_0^{(\hat{\alpha})} = 0, \tag{9}$$

where $\text{sign}(\cdot)$ is defined in (10). Use of (9) enables computation of a data-fitting Markov chain for $j = 1, \ldots, n$, where $n\Delta t = T$ is the size of the window.

To illustrate, we used the parameter estimates[30] $\hat{\mu} = 0.0256$ and $\hat{\sigma} = 0.1486$ obtained for 08/24/2021 to generate the pricing tree via (7) and (9) over the period 08/25/2020 to 08/24/2021.[31]

---

[30] With $r_{j\Delta t}^{(\hat{\alpha}, \hat{\mu}, \hat{\sigma}, n)}$ in (8) replaced by $r_{j\Delta t}^{(\text{emp})}$ in (9) and with $M_0^{(\hat{\alpha})} = 0$, the pricing tree no longer has explicit dependence on $\alpha$ (only implicitly through the empirical return).

[31] We use the price of SPY on 08/24/2020 as $S_0^{(\text{emp})}$.

The resulting price series $S^{(\hat{\mu},\hat{\sigma},n)}$ is shown in Fig. 2, where it is compared to that of the empirical SPY price, $S^{(SPY)}$. The R-squared value of the fit is 0.94, the RMSE is 10.2.

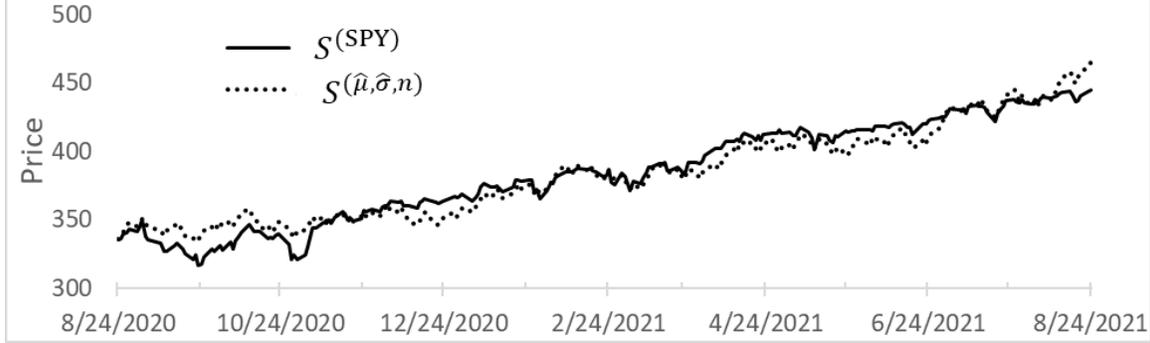

**Figure 2.** The fit of the model price $S^{(\hat{\mu},\hat{\sigma},n)}$ to the SPY market price $S^{(SPY)}$ over the time period 8/25/2020 through 8/24/2021

### 4.2. Stochastic dynamics of Itô-McKean skew Brownian motion

The goal in this subsection and the next is to motivate the market complete tree model which we will introduce in Section 5.[32] The stochastic differential equation for $\mathbb{A}^{(\delta)}$ is a derived directly from Tanaka's formula[33]. If $\mathbb{B} = \{B_t,\ t \geq 0\}$ is a Brownian motion, then

$$|B_t| = \int_0^t \text{sign}(B_s)dB_s + L_t^{(0,\mathbb{B})}, \quad \text{where } \text{sign}(x) = \begin{cases} 1, & \text{if } x > 0, \\ 0, & \text{if } x = 0, \\ -1, & \text{if } x < 0. \end{cases} \quad (10)$$

In (10), $\mathbb{L}^{(a,\mathbb{B})} = \{L_t^{(a,\mathbb{B})}, t \geq 0\}$ is the Brownian motion local time at $a \in R$ defined by

$$L_t^{(a,\mathbb{B})} = \lim_{\varepsilon \downarrow 0} \frac{1}{2\varepsilon} \int_0^t I_{\{a-\varepsilon \leq B_s \leq a+\varepsilon\}} ds = \lim_{\varepsilon \downarrow 0} \frac{1}{2\varepsilon} \text{Leb}\{s \in [0,t]: B_s \in (a-\varepsilon, a+\varepsilon)\}, \quad (11)$$

where Leb$\{\cdot\}$ is the Lebesgue measure. Informally,[34] $L_t^{(a;\mathbb{B})} = \int_0^t \delta_a(B_s)ds$, where $\delta_a(x), x \in R$, is the Dirac delta function at $a$, and

$$dL_t^{(a,\mathbb{B})} = \delta_a(B_t)dt. \tag{12}$$

---

[32] Subsections 4.2 and 4.3 are of illustrative (informal) character. The rigorous results are presented in the cited references.
[33] See Chung and Williams (1991, Section 7.3) and Björk (2019, Chapter 4).
[34] Björk (2019, Proposition 2.4).

For $a \in R$, $L_\infty^{(a;\mathbb{B})} = \infty$ with probability $1$[35]. If $\mathbb{M}^{(\mathbb{B})} = \left\{ m_t^{(\mathbb{B})} = \max_{0 \leq s \leq t} B_s, \ t \geq 0 \right\}$ and $\mathbb{L}^{(0;\mathbb{B})} = \left\{ L_t^{(0;\mathbb{B})}, t \geq 0 \right\}$, then $\mathbb{M}^{(\mathbb{B})} \stackrel{d}{=} \mathbb{L}^{(0;\mathbb{B})}$.[36] From (10), (11) and (12),

$$dA_t^{(\delta)} = \sqrt{1-\delta^2}dB_{1,t} + \delta d|B_{2,t}| = dB_{1\leftrightarrow 2,t} + \delta dL_t^{(0,\mathbb{B}_2)}, \tag{13}$$

where $\mathbb{B}_{1\leftrightarrow 2,t} = \{B_{1\leftrightarrow 2,t} = \sqrt{1-\delta^2}B_{1,t} + \delta\,\text{sign}(B_{2,t})B_{2,t}, \ t \geq 0\}$ is a standard BM.

### 4.3 The risk-free rate in a market with uncertainties determined by Itô-Mckean SBM

We now extend the Black-Scholes-Merton market model[37] assuming the following market model $(\mathcal{S}, \mathcal{B}, \mathcal{C})$, with $\mathcal{S}, \mathcal{B}$ and $\mathcal{C}$ defined as follows.

$(i)$ $\mathcal{S}$ is a risky asset (stock) with price dynamics $S_t, t \in [0,T], 0 < T < \infty$ given by geometric Itô-Mckean SBM,

$$S_t = S_0 \exp\left(\mu t + \sigma A_t^{(\delta)}\right), \ S_0 > 0, \ \mu \in R, \ \sigma > 0. \tag{14}$$

In (14), $\mathbb{A}^{(\delta)} = \left\{ A_t^{(\delta)}, t \geq 0 \right\}$ is an Itô-Mckean SBM as defined in (1), with dynamics determined by (13), where $\mathbb{B}_1 = \{B_{1,t}, \ t \geq 0\}$, and $\mathbb{B}_2 = \{B_{2,t}, \ t \geq 0\}$ are two independent Brownian motions generating a stochastic basis $\left(\Omega, \mathbb{F} = \left\{\mathcal{F}_t = \sigma\left((B_{1,s}, B_{2,s}), 0 \leq s \leq t\right), t \geq 0\right\}, \mathbb{P}\right)$.

$(ii)$ $\mathcal{B}$ is a riskless asset (bond) with price dynamics

$$\beta_t = \beta_0 e^{rt}, t \in [0,T], \ \beta_0 > 0, \ r \geq 0. \tag{15}$$

---

[35] Revuz and Yor (1999, Corollary 2.4, Chapter VI)
[36] A more general result, known as the Lévy Theorem, is true:
$$\left\{(m_t^{(\mathbb{B})} - B_t, m_t^{(\mathbb{B})}), t \geq 0\right\} \stackrel{d}{=} \left\{(|B_t|, \mathbb{L}_t^{(0;\mathbb{B})}), t \geq 0\right\}.$$
See Revuz and Yor (1999, Theorem 2.3, Chapter VI).
[37] See Black and Scholes (1973), Merton (1973), and Duffie (2001, Chapters 5 and 6).

(iii) $\mathcal{C}$ is a European Contingent Claim (ECC)[38] with underlying asset $\mathcal{S}$ and maturity time $T$. The price dynamics of $\mathcal{C}$ is given by

$$f_t = f(S_t, t), t \in [0, T), \quad f_T = g(S_T),$$

where $f(x, t) \in R, x > 0, t \in [0, T)$ and $g(x), x > 0$ satisfy the usual regularity conditions[39].

In the market model $(\mathcal{S}, \mathcal{B}, \mathcal{C})$ where the underlying risky asset $\mathcal{S}$ has price dynamics (14), the riskless rate $r \geq 0$ in (15) cannot be arbitrarily chosen. This is due to the delta-function jump in the instantaneous mean return of the stock; that is, from (13) and (12), $dL_t^{(0)}(\mathbb{B}_2) = \delta_0(B_{2,t})dt$.[40] To illustrate the presence of the Dirac delta function term in the riskless rate, consider a market model with two risky assets, $\mathcal{S}' = (\mathcal{S}^{(1)}, \mathcal{S}^{(2)})$[41], each having price dynamics

$$S_t^{(i)} = S_0^{(i)} \exp\left(\mu^{(i)} t + \sigma^{(i)} A_t^{(\delta^{(i)})}\right), \quad S_0^{(i)} > 0, \quad \mu^{(i)} \in R, \quad \sigma^{(i)} > 0, \quad i = 1,2, \tag{16}$$

where

$$A_t^{(\delta^{(i)})} = \sqrt{1 - \delta^{(i)^2}} B_{1,t} + \delta^{(i)} |B_{2,t}| = B_{1 \leftrightarrow 2,t} + \delta^{(i)} L_t^{(0)}(\mathbb{B}_2), \quad \delta^{(i)} \in (-1,1). \tag{17}$$

From (16), (17) and the extended Itô formula[42], it follows that the dynamics of $S_t^{(i)}$, $t \geq 0$, $i = 1,2$, is determined by

$$dS_t^{(i)} = S_t^{(i)} \left\{\left(\mu^{(i)} + \frac{1}{2}\sigma^{(i)^2} + \sigma^{(i)}\delta^{(i)}\delta_0(B_{2,t})\right) dt + \sigma^{(i)} dB_{1 \leftrightarrow 2,t}\right\}.$$

To evaluate the riskless rate determined by the market model $\mathcal{S}^{(\text{IM})}$, we follow the approach in Section 6D in Duffie (2001)[43]. The market $(\mathcal{S}', \mathcal{B}, \mathcal{C})$ is free of arbitrage and complete if there exists a unique state price deflator $\Pi = \{\pi_t, t \geq 0\}$ on $\left(\Omega, \mathbb{F} = \left\{\mathcal{F}_t = \sigma\left((B_{1,s}, B_{2,s}), 0 \leq s \leq t\right), t \geq 0\right\}, \mathbb{P}\right)$. We search for $\Pi$ having dynamics

$$d\pi_t = \pi_t\left(\mu_t^{(\pi)} dt + \sigma_t^{(\pi)} dB_{1\leftrightarrow 2,t}\right),$$

such that $S_t^{(\pi,i)} = S_t^{(i)} \pi_t$, $i = 1,2$ are martingales on $(\Omega, \mathbb{F}, \mathbb{P})$. The martingale condition implies that

$$\mu_t^{(\pi)} = \frac{\left(\mu^{(1)} + \frac{1}{2}\sigma^{(1)^2} + \sigma^{(1)}\delta^{(1)}\delta_0(B_{2,t})\right)\sigma^{(2)} - \left(\mu^{(2)} + \frac{1}{2}\sigma^{(2)^2} + \sigma^{(2)}\delta^{(2)}\delta_0(B_{2,t})\right)\sigma^{(1)}}{\sigma^{(1)} - \sigma^{(2)}},$$

and

$$\sigma_t^{(\pi)} = \frac{\mu^{(1)} + \frac{1}{2}\sigma^{(1)^2} + \sigma^{(1)}\delta^{(1)}\delta_0(B_{2,t}) - \left(\mu^{(2)} + \frac{1}{2}\sigma^{(2)^2} + \sigma^{(2)}\delta^{(2)}\delta_0(B_{2,t})\right)}{\sigma^{(1)} - \sigma^{(2)}}.$$

---

[38] We use terms ECC, options, and derivative interchangeably.
[39] Duffie (2001, Chapter 5).
[40] Björk (2019, Theorem 4.1).
[41] Heuristically, one can view $\mathcal{S}^{(1)}$ and $\mathcal{S}^{(2)}$ as two portfolios on the "capital market line".
[42] Björk (2019, Theorem 4.3).
[43] The general framework of a market without a risk-free asset was first outlined by Black (1972) in his development of the zero-beta capital asset pricing model.

The riskless rate $r'_t$ in the market $(\mathcal{S}', \mathcal{B}, \mathcal{C})$ is now determined by $\pi_t$ via[44]

$$r'_t = -\mu_t^{(\pi)} = \frac{\mu^{(2)}\sigma^{(1)} - \mu^{(1)}\sigma^{(2)} + \frac{1}{2}\left(\sigma^{(2)^2} - \sigma^{(1)^2}\right) + \sigma^{(1)}\sigma^{(2)}\left(\delta^{(2)} - \delta^{(1)}\right)\delta_0(B_{2,t})}{\sigma^{(1)} - \sigma^{(2)}}.$$

Thus, as long as

$$\delta^{(1)} = \delta^{(2)} = \delta \in (-1,1) \tag{18}$$

then[45]

$$r'_t = r^{(FB)} = \frac{\mu^{(2)}\sigma^{(1)} - \mu^{(1)}\sigma^{(2)} + \frac{1}{2}\left(\sigma^{(2)^2} - \sigma^{(1)^2}\right)}{\sigma^{(1)} - \sigma^{(2)}}. \tag{19}$$

To guarantee no arbitrage, we shall assume in Section 5, that (18) holds.

---

[44] Duffie (2001, Section 6D).
[45] We adopt the notation $r^{(FB)}$ in (19) to acknowledge that it follows ideas from Black (1972) for riskless rate computation. See footnote 38.

## 5. Derivative pricing with an Itô-Mckean skew pricing tree

While the market model $(\mathcal{S}', \mathcal{B}, \mathcal{C})$ in Section 4.3 is arbitrage-free and complete, as the price uncertainty is determined by $\mathbb{A}^{(\delta)}$, any derivative price $f_t, t \in [0, T]$ is affected by three market uncertainties $\left(\mathbb{B}_1, \mathbb{B}_2, \mathbb{L}^{(0,\mathbb{B}_2)}\right)$. While the stochastic drift $\mathbb{L}^{(0,\mathbb{B}_2)}$ is completely determined by $\mathbb{B}_2$, it enters the deterministic drift in the asset price as a Dirac delta function term. This term must be removed when a hedging portfolio is constructed. To remove this term while retaining market completeness, we shall assume that

$$f_t = f(S_t, t), \qquad S_t = \left(S_t^{(1)}, S_t^{(2)}, S_t^{(3)}\right), \quad t \in [0, T], \tag{20}$$

where the triplet of underlying asset prices, $S_t$, has price dynamics determined by

$$S_t^{(i)} = S_0^{(i)} \exp\left(\mu^{(i)} t + \sigma^{(i)} A_t^{(\delta)}\right), \quad \mu^{(i)} \in R, \quad \sigma^{(i)} \in R, \quad \delta \in (-1,1), \quad i = 1,2,3. \tag{21}$$

We have chosen one and the same skewness parameter $\delta \in (-1,1)$ for the three price processes following our conclusion in Section 4,3; see (18).

### 5.1. Completeness of the IMSPT

We show that the market model $\left(\mathcal{S} = \left(\mathcal{S}^{(1)}, \mathcal{S}^{(2)}, \mathcal{S}^{(3)}\right), \mathcal{B}, \mathcal{C}\right)$ determined by (21), (15) and (20) is complete. First, we will derive the price dynamics of $S_t, t \in [0, T]$, as a limit of two dependent binomial trees. Let $\xi_k^{(j)}$, $k \in \mathcal{N}$, $j = 1, 2$, with $\xi_0^{(j)} = 0$, be a sequence of pairs of independent random signs, $\mathbb{P}\left(\xi_k^{(j)} = 1\right) = 1 - \mathbb{P}\left(\xi_k^{(j)} = -1\right) = \frac{1}{2}$.[46] Let $\Delta t = T/n$ and $\zeta_k^{(j,n)} = \sum_{i=1}^{k} \xi_i^{(j)}$, $k = 1, \ldots, n$, $\zeta_0^{(j,n)} = 0$. Consider the $\mathfrak{D}[0,T]$ processes

$$\mathbb{B}_j^{(n)} = \begin{cases} B_{j,t}^{(n)} = \sqrt{\Delta t} \zeta_k^{(j,n)} & \text{for } t \in [k\Delta t, (k+1)\Delta t), \quad k = 0, \ldots, n-1, \\ B_{j,T}^{(n)} = \sqrt{\Delta t} \zeta_n^{(j,n)}, \end{cases}, \quad j = 1, 2.$$

---

[46] The sequences $\xi_k^{(1)}$ and $\xi_k^{(2)}, k \in \mathcal{N}$, are assumed independent.

Then, as $n \uparrow \infty$, $\mathbb{B}_j^{(n)}$ converges weakly in $\mathcal{D}[0,T]$ to $\mathbb{B}_i$ by DPIP. Let

$$\mathbb{A}^{(\delta,n)} = \left\{ A_t^{(\delta,n)} = \sqrt{1-\delta^2} B_{1,t}^{(n)} + \delta \left| B_{2,t}^{(n)} \right|, \ t \in [0,T] \right\}.$$

Then $\mathbb{A}^{(\delta,n)}$ converges weakly in $\mathcal{D}[0,T]$ to $\mathbb{A}^{(\delta)}$. Define the filtration $\mathbb{F}^{(\xi,n)} = \left\{ \mathcal{F}_k^{(\xi,n)} = \sigma\left[ \left(\xi_0^{(1)}, \xi_0^{(2)}\right), \ldots, \left(\xi_k^{(1)}, \xi_k^{(2)}\right) \right], k = 0, \ldots, n \right\}$. For $k = 0, \ldots, n-1$, conditionally on $\mathcal{F}_k^{(\xi,n)}$,

$$A_{(k+1)\Delta t}^{(\delta,n)} - A_{k\Delta t}^{(\delta,n)} = \begin{cases} \left[ \sqrt{1-\delta^2} + \delta \left( \left|\zeta_k^{(2,n)} + 1\right| - \left|\zeta_k^{(2,n)}\right| \right) \right] \sqrt{\Delta t} & \text{w.p. } 1/4, \\ \left[ \sqrt{1-\delta^2} + \delta \left( \left|\zeta_k^{(2,n)} - 1\right| - \left|\zeta_k^{(2,n)}\right| \right) \right] \sqrt{\Delta t} & \text{w.p. } 1/4, \\ \left[ -\sqrt{1-\delta^2} + \delta \left( \left|\zeta_k^{(2,n)} + 1\right| - \left|\zeta_k^{(2,n)}\right| \right) \right] \sqrt{\Delta t} & \text{w.p. } 1/4, \\ \left[ -\sqrt{1-\delta^2} + \delta \left( \left|\zeta_k^{(2,n)} - 1\right| - \left|\zeta_k^{(2,n)}\right| \right) \right] \sqrt{\Delta t} & \text{w.p. } 1/4. \end{cases}$$

Consider the triplet of dependent binomial pricing trees $S_{k\Delta t}^{(n)} = \left( S_{k\Delta t}^{(1,n)}, S_{k\Delta t}^{(2,n)}, S_{k\Delta t}^{(3,n)} \right)$, $k = 0, \ldots, n$, where

$$S_{k\Delta t}^{(i,n)} = S_0^{(i)} \exp\left( \mu^{(i)} k \Delta t + \sigma^{(i)} A_{k\Delta t}^{(\delta,n)} \right), \ i = 1,2,3.$$

Conditionally on $\mathcal{F}_k^{(\xi,n)}$,

$$S_{(k+1)\Delta t}^{(i,n)} = S_{k\Delta t}^{(i,n)} \exp\left\{ \mu^{(i)} \Delta t + \sigma^{(i)} \left( A_{(k+1)\Delta t}^{(\delta,n)} - A_{k\Delta t}^{(\delta,n)} \right) \right\}$$

$$= \begin{cases} S_{(k+1)\Delta t}^{(i,n;uu)} = \psi_{k+1}^{(i,uu,n)} & \text{w.p. } 1/4, \\ S_{(k+1)\Delta t}^{(i,n;ud)} = \psi_{k+1}^{(i,uu,n)} & \text{w.p. } 1/4, \\ S_{(k+1)\Delta t}^{(i,n;du)} = \psi_{k+1}^{(i,du,n)} & \text{w.p. } 1/4, \\ S_{(k+1)\Delta t}^{(i,n;dd)} = \psi_{k+1}^{(i,dd,n)} & \text{w.p. } 1/4, \end{cases} \qquad (22)$$

where

$$\psi_{k+1}^{(i,uu,n)} = S_{k\Delta t}^{(i,n)} \exp\left[ \mu^{(i)} \Delta t + \sigma^{(i)} \left\{ \sqrt{1-\delta^2} + \delta \left( \left|\zeta_k^{(2,n)} + 1\right| - \left|\zeta_k^{(2,n)}\right| \right) \right\} \sqrt{\Delta t} \right],$$

$$\psi_{k+1}^{(i,ud,n)} = S_{k\Delta t}^{(i,n)} \exp\left[ \mu^{(i)} \Delta t + \sigma^{(i)} \left\{ \sqrt{1-\delta^2} + \delta \left( \left|\zeta_k^{(2,n)} - 1\right| - \left|\zeta_k^{(2,n)}\right| \right) \right\} \sqrt{\Delta t} \right],$$

$$\psi_{k+1}^{(i,du,n)} = S_{k\Delta t}^{(i,n)} \exp\left[ \mu^{(i)} \Delta t + \sigma^{(i)} \left\{ -\sqrt{1-\delta^2} + \delta \left( \left|\zeta_k^{(2,n)} + 1\right| - \left|\zeta_k^{(2,n)}\right| \right) \right\} \sqrt{\Delta t} \right],$$

$$\psi_{k+1}^{(i,dd,n)} = S_{k\Delta t}^{(i,n)} \exp\left[\mu^{(i)}\Delta t + \sigma^{(i)}\left\{-\sqrt{1-\delta^2} + \delta\left(\left|\zeta_k^{(2,n)} - 1\right| - \left|\zeta_k^{(2,n)}\right|\right)\right\}\sqrt{\Delta t}\right],$$

for $k = 0, \ldots, n$, $i = 1,2,3$. We call (22) the *Itô-McKean skew pricing tree*. It generates a $\mathfrak{D}[0,T]^{\times 3}$- valued process $\mathbb{S}^{(n;T)} = \left\{\left(S_t^{(1,n)}, S_t^{(2,n)}, S_t^{(3,n)}\right), t \in [0,T]\right\}$, where

$$S_t^{(i,n)} = \left\{S_{k\Delta t}^{(i,n)} \text{ for } t \in [k\Delta t, (k+1)\Delta t), k = 0, \ldots, n-1, S_T^{(i,n)} = S_{n\Delta t}^{(i,n)}\right\}, i = 1,2,3.$$

By the DPIP, $\mathbb{S}^{(n;T)}$ weakly converges to $\mathbb{S}^{(T)} = \{S_t, t \in [0,T]\}$ as $n \uparrow \infty$.

**Remark 3:** In the special case when the second random sign sequence $\xi_k^{(2)} = 0$, $k \in \mathcal{N}$, the SRW tree (22) becomes a single binomial tree with up and down prices given, respectively, by $S_{(k+1)\Delta t}^{(i,n,u)} = S_{k\Delta t}^{(i,n)} \exp\{\mu^{(i)}\Delta t + \sigma^{(i)}(\sqrt{1-\delta^2} + \delta)\sqrt{\Delta t}\}$ and $S_{(k+1)\Delta t}^{(i,n,d)} = S_{k\Delta t}^{(i,n)} \exp\{\mu^{(i)}\Delta t + \sigma^{(i)}(-\sqrt{1-\delta^2} + \delta)\sqrt{\Delta t}\}$. The upward risk-neutral probability then becomes $q_{(k+1)\Delta t}^{(n;u)} = \left(1 - \theta\sqrt{\Delta t}\right)/2$, where $\theta = (\bar{\mu} - \bar{\sigma}^2/2 - r)/\bar{\sigma}$.

### 5.2. Option pricing with the IMSPT

Suppose the ECC $\mathcal{C}$, with the triplet of underlying assets $S_t$, $t \in [0,T]$ and maturity time $T$, has a price process $f_t = f(S_t, t)$, $t \in [0,T)$, where $f(x,t)$ is a sufficiently smooth, real-valued function on $(0,\infty)^3 \times [0,T)$. The payoff of $\mathcal{C}$ at maturity is $g(S_T)$ for some real-valued function $g(x), x \in (0,\infty)^3$.[47] Consider the finite price dynamics $f_{k\Delta t}^{(n)} = f\left(S_{k\Delta t}^{(n)}, k\Delta t\right)$ of $\mathcal{C}$ on the lattice $k = 0, \ldots, n$. Conditionally on $\mathcal{F}_k^{(\xi,n)}$,

$$f_{(k+1)\Delta t}^{(n)} = \begin{cases} f_{(k+1)\Delta t}^{(n;uu)} = f\left(S_{(k+1)\Delta t}^{(n;uu)}, (k+1)\Delta t\right), \\ f_{(k+1)\Delta t}^{(n;ud)} = f\left(S_{(k+1)\Delta t}^{(n;ud)}, (k+1)\Delta t\right), \\ f_{(k+1)\Delta t}^{(n;du)} = f\left(S_{(k+1)\Delta t}^{(n;du)}, (k+1)\Delta t\right), \\ f_{(k+1)\Delta t}^{(n;dd)} = f\left(S_{(k+1)\Delta t}^{(n;dd)}, (k+1)\Delta t\right), \end{cases}$$

---

[47] $f$ and $g$ satisfy the usual regularity conditions (Duffie 2001, Chapters 5 and 6).

where $S^{(n;uu)}_{(k+1)\Delta t} = \left(S^{(1,n;uu)}_{(k+1)\Delta t}, S^{(2,n;uu)}_{(k+1)\Delta t}, S^{(3,n;uu)}_{(k+1)\Delta t}\right)$, with analogous definitions for $S^{(n;ud)}_{(k+1)\Delta t}$, $S^{(n;du)}_{(k+1)\Delta t}$, and $S^{(n;dd)}_{(k+1)\Delta t}$. For $k = 0, 1, \ldots, n-1$, consider the replicating risk-neutral portfolio $P^{(n)}_{k\Delta t} = \sum_{i=1}^{3} D^{(i,n)}_{k\Delta t} S^{(i,n)}_{k\Delta t} - f^{(n)}_{k\Delta t}$, where $D^{(i,n)}_{k\Delta t}$, $i = 1,2,3$ are the hedging deltas to be determined later on. Then, conditionally on $\mathcal{F}^{(\xi,n)}_k$, $P^{(n)}_{(k+1)\Delta t} = \sum_{i=1}^{3} D^{(i,n)}_{k\Delta t} S^{(i,n)}_{(k+1)\Delta t} - f^{(n)}_{(k+1)\Delta t}$. From the risk-neutrality condition, it follows that

$$P^{(n,uu)}_{(k+1)\Delta t} = P^{(n,ud)}_{(k+1)\Delta t} = P^{(n,du)}_{(k+1)\Delta t} = P^{(n,dd)}_{(k+1)\Delta t}, \qquad (23)$$

where

$$P^{(n,uu)}_{(k+1)\Delta t} = \sum_{i=1}^{3} D^{(i,n)}_{k\Delta t} S^{(i,n;uu)}_{(k+1)\Delta t} - f^{(n;uu)}_{(k+1)\Delta t}, \quad P^{(n,ud)}_{(k+1)\Delta t} = \sum_{i=1}^{3} D^{(i,n)}_{k\Delta t} S^{(i,n;ud)}_{(k+1)\Delta t} - f^{(n;ud)}_{(k+1)\Delta t},$$
$$P^{(n,du)}_{(k+1)\Delta t} = \sum_{i=1}^{3} D^{(i,n)}_{k\Delta t} S^{(i,n;du)}_{(k+1)\Delta t} - f^{(n;du)}_{(k+1)\Delta t}, \quad P^{(n,dd)}_{(k+1)\Delta t} = \sum_{i=1}^{3} D^{(i,n)}_{k\Delta t} S^{(i,n;dd)}_{(k+1)\Delta t} - f^{(n;dd)}_{(k+1)\Delta t}. \qquad (24)$$

Then (23) and (24) imply that the hedging deltas $D^{(i,n)}_{k\Delta t}$, $i = 1,2,3$, are given by

$$D^{(i,n)}_{k\Delta t} = \frac{\mathbb{D}^{(i)}_{k\Delta t}}{\mathbb{D}_{k\Delta t}}, \quad i = 1, 2, 3, \qquad (25)$$

where the 3x3 determinants $\mathbb{D}_{k\Delta t}$, $\mathbb{D}^{(i)}_{k\Delta t}$, $i = 1, 2, 3$ are defined in Appendix D. Given $\mathcal{F}^{(\xi,n)}_k$, portfolio $P^{(n)}_{(k+1)\Delta t} = \sum_{i=1}^{3} D^{(i,n)}_{k\Delta t} S^{(i,n)}_{(k+1)\Delta t} - f^{(n)}_{(k+1)\Delta t}$ is riskless. This leads to

$$P^{(n)}_{k\Delta t} = \sum_{i=1}^{3} D^{(i,n)}_{k\Delta t} S^{(i,n)}_{k\Delta t} - f^{(n)}_{k\Delta t} = e^{-r\Delta t} P^{(n,uu)}_{(k+1)\Delta t} = e^{-r\Delta t} \sum_{i=1}^{3} D^{(i,n)}_{k\Delta t} S^{(i,n;uu)}_{(k+1)\Delta t} - f^{(n;uu)}_{(k+1)\Delta t}.$$

Thus, given $\mathcal{F}^{(\xi,n)}_k$, the derivative price $f^{(n)}_{k\Delta t}$ at $k\Delta t$ is given by

$$f^{(n)}_{k\Delta t} = \sum_{i=1}^{3} D^{(i,n)}_{k\Delta t} S^{(i,n)}_{k\Delta t} - e^{-r\Delta t} \sum_{i=1}^{3} D^{(i,n)}_{k\Delta t} S^{(i,n;uu)}_{(k+1)\Delta t} + f^{(n;uu)}_{(k+1)\Delta t}. \qquad (26)$$

Now, (25) and (26) imply that

$$f_{k\Delta t}^{(n)} = e^{-r\Delta t}\left(q_{(k+1)\Delta t}^{(n;uu)}f_{(k+1)\Delta t}^{(n;uu)} + q_{(k+1)\Delta t}^{(n;ud)}f_{(k+1)\Delta t}^{(n;ud)} + q_{(k+1)\Delta t}^{(n;du)}f_{(k+1)\Delta t}^{(n;du)} + q_{(k+1)\Delta t}^{(n;du)}f_{(k+1)\Delta t}^{(n;du)}\right). \quad (27)$$

In (27) the conditional-on-$\mathcal{F}_k^{(\xi,n)}$ risk-neutral probabilities $q_{(k+1)\Delta t}^{(n;uu)}$, $q_{(k+1)\Delta t}^{(n;ud)}$, $q_{(k+1)\Delta t}^{(n;du)}$, and $q_{(k+1)\Delta t}^{(n;du)}$, have the form,

$$q_{(k+1)\Delta t}^{(n;uu)} = \frac{\mathbb{Q}_{(k+1)\Delta t}^{(n;uu)}}{\mathbb{Q}_{(k+1)\Delta t}}, \quad q_{(k+1)\Delta t}^{(n;ud)} = \frac{\mathbb{Q}_{(k+1)\Delta t}^{(n;ud)}}{\mathbb{Q}_{(k+1)\Delta t}}, \quad q_{(k+1)\Delta t}^{(n;du)} = \frac{\mathbb{Q}_{(k+1)\Delta t}^{(n;du)}}{\mathbb{Q}_{(k+1)\Delta t}}, \quad q_{(k+1)\Delta t}^{(n;dd)} = \frac{\mathbb{Q}_{(k+1)\Delta t}^{(n;dd)}}{\mathbb{Q}_{(k+1)\Delta t}}, \quad (28)$$

where the terms in (28) are defined in Appendix D.

**Remark 4:** Consider the special case $\delta = 0$. Then the underlying price process follows the Jarrow-Rudd binomial tree[48],

$$S_{(k+1)\Delta t}^{(1,n)} = S_{k\Delta t}^{(1,n)}\begin{cases}\exp(\mu^{(1)}\Delta t + \sigma^{(1)}\sqrt{\Delta t}) & \text{if } \xi_k^{(1)} = 1, \\ \exp(\mu^{(1)}\Delta t - \sigma^{(1)}\sqrt{\Delta t}) & \text{if } \xi_k^{(1)} = -1,\end{cases}$$

having the risk neutral probabilities

$$q_{(k+1)\Delta t}^{(n;u)} = \frac{e^{(r-\mu^{(1)})\Delta t} - e^{-\sigma^{(1)}\sqrt{\Delta t}}}{e^{\sigma^{(1)}\sqrt{\Delta t}} - e^{-\sigma^{(1)}\sqrt{\Delta t}}}, \qquad q_{(k+1)\Delta t}^{(n;d)} = 1 - q_{(k+1)\Delta t}^{(n;u)}.$$

With $o(\Delta t) = 0$, $q_{(k+1)\Delta t}^{(n;u)} = \frac{1}{2} - \frac{1}{2}\theta^{(1)}\sqrt{\Delta t}$, where $\theta^{(1)} = \frac{\mu^{(1)} - \sigma^{(1)^2}/2 - r}{\sigma^{(1)}}$ is the market price of risk (Kim et al., 2016, 2019).

**Remark 5:** Consider the special case $\delta = 1$. Then the underlying price process follows the path-dependent binomial tree; conditionally on $\mathcal{F}_k^{(\xi^{(2)},n)} = \sigma\left(\xi_0^{(2)}, \dots, \xi_k^{(2)}\right)$,

$$S_{(k+1)\Delta t}^{(2,n)} = S_{k\Delta t}^{(2,n)}\begin{cases}\exp\left(\mu^{(2)}\Delta t + \sigma^{(2)}\left(\left|\zeta_k^{(2,n)} - 1\right| - \left|\zeta_k^{(2,n)}\right|\right)\sqrt{\Delta t}\right) & \text{w.p. } 1/2, \\ \exp\left(\mu^{(2)}\Delta t + \sigma^{(2)}\left(\left|\zeta_k^{(2,n)} + 1\right| - \left|\zeta_k^{(2,n)}\right|\right)\sqrt{\Delta t}\right) & \text{w.p. } 1/2,\end{cases}$$

for $k = 0, 1, \dots, n - 1$. In this case, the risk the risk neutral probabilities are

$$q_{(k+1)\Delta t}^{(n;u)} = \frac{e^{(r-\mu^{(2)})\Delta t} - e^{\sigma^{(2)}\left(\left|\zeta_k^{(2,n)}-1\right|-\left|\zeta_k^{(2,n)}\right|\right)\sqrt{\Delta t}}}{e^{\sigma^{(2)}\left(\left|\zeta_k^{(2,n)}+1\right|-\left|\zeta_k^{(2,n)}\right|\right)\sqrt{\Delta t}} - e^{\sigma^{(2)}\left(\left|\zeta_k^{(2,n)}-1\right|-\left|\zeta_k^{(2,n)}\right|\right)\sqrt{\Delta t}}}, \qquad q_{(k+1)\Delta t}^{(n;d)} = 1 - q_{(k+1)\Delta t}^{(n;u)}$$

given $\mathcal{F}_k^{(\xi^{(2)},n)} = \sigma\left(\xi_0^{(2)}, \ldots, \xi_k^{(2)}\right)$, $k = 0, \ldots, n-1$.

---

[48] See Jarrow and Rudd (1983, Chapters V & VI), Hull (2012, p. 442), and Kim et al. (2016, 2019).

### 5.3. Numerical example - option pricing

We apply the results developed in Section 5.2. Based on our conclusion in Section 4.3 (embodied in equation (18)), we assume the same skewness parameter $\delta$ for all three risky assets. To ensure the accuracy of this assumption, we choose three ETFs that track sub-indices of the same composite index. Specifically, we consider the S&P Global 1200 index (SPG1200) and the ETFs: SPY, iShares Europe ETF (IEV) and iShares JPX-NIKKEI 400 ETF (JPXN) which track the three major components[49] of this index.

There are strong numerical considerations behind our choice of three ETFs tracking sub-indices of a composite index. Consider an alternate choice consisting of the S&P 500 index and the three ETFs: SPY, IVV (iShares Core S&P 500 ETF); and VOO (Vanguard S&P 500 ETF) each of which tracks the full index. The similarity of the price processes governing these three ETFs leads to numerical instability (division by very small denominators) in the computations of the risk neutral probabilities (28). That this is so can be seen more immediately in the case of two-risky assets discussed in Section 4.3. Computation of the riskless rate (28) diverges whenever $\sigma^{(1)} - \sigma^{(2)} \downarrow 0$, which happens frequently when the ETFs closely track the same index. By choosing ETFs that track separate components of a composite index, we ensure that the volatilities of each ETF have both a systemic and an idiosyncratic component. In practice, the (differing) systemic components will ensure that the values of the required denominators remain sufficiently large compared to the value of their numerator terms.

---

[49] S&P Global 1200 is a composite index comprised of seven regional and country indices. The major three are the S&P 500, S&P Europe 350, and S&P/TOPIX 150 (Japan).

We compute $\delta$ assuming that the dynamics of the SPG1200 index follows the GSBM

$$S_t^{(SPG)} = S_0^{(SPG)} \exp\left(\mu^{(SPG)}t + \sigma^{(SPG)}A_t^{(\delta^{(SPG)})}\right), \quad t \in [0,T], \ \mu^{(SPG)} \in R, \ \sigma^{(SPG)} > 0, \quad (29)$$

where $A_t^{(\delta^{(SPG)})}$ is given by (1). As noted in section 3, $\mathbb{A}^{(\delta)} \stackrel{d}{=} \mathbb{B}^{(\alpha=(1+\delta)/2)}$ for $\delta \in (-1,1)$. As the processes on a time interval $[0,T]$ defined by either (29) or (6) generate the same probability law (Corns and Satchell 2007), we use the procedure outlined in Section 4.1 on the SPG1200 index values over the period 09/13/2016 through 09/14/2021 using 1-year moving windows to generate a time series $\hat{\delta}_t^{(SPG)} = 2\hat{\alpha}_t^{(SPG,Med)} - 1$, where, as above, "Med" refers to a moving 1-month median value. As in Section 4.1, we restrict values of $\hat{\alpha}$ to the range [0.45,0.55]. The resultant time series, $\hat{\delta}_t^{(SPG)}$, is shown in Fig. 3.

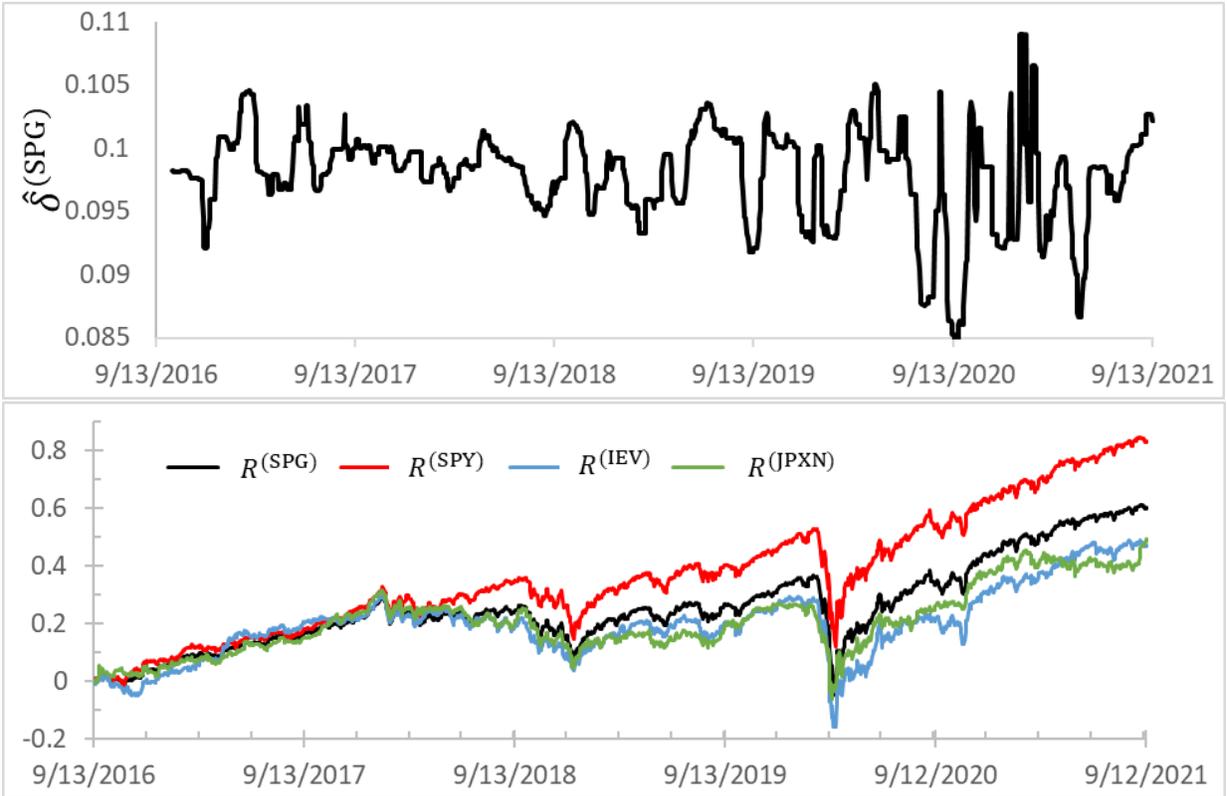

**Figure 3.** (top) Time series of $\hat{\delta}^{(SPG)}$ estimated for the S&P Global 1200 index over the period 09/13/2016 through 09/14/2021. (bottom) Cumulative return time series of the three ETFs, SPY, IEV, and JPXN, as well as that of the SPG1200 index (SPG) over the same period.

We model the price dynamics of each ETF over the one-year (252 trading day) period 9/15/2020 - 9/14/2021 by

$$S_t^{(i)} = S_0^{(i)} \exp\left(\mu^{(i)} t + \sigma^{(i)} A_t^{(\delta^{(SPG)})}\right), \quad i = \text{SPY, IEV, JPXN,}$$

where $\delta^{(SPG)} = 0.102$ is the value obtained for 09/14/2021. To compute the values $\mu^{(i)}$ and $\sigma^{(i)}$, we perform least-squares minimization on the relation

$$R_{k\Delta t}^{(i;\text{emp})} - \mu^{(i)} k\Delta t - \sigma^{(i)} \delta^{(SPG)} \sqrt{\frac{2k\Delta t}{\pi}} = \varepsilon_{k\Delta t}^{(i)}. \tag{30}$$

Table 2 presents the fitted estimates $\hat{\mu}^{(i)}$ and $\hat{\sigma}^{(i)}, i = \text{SPY, IEV, JPXN}$, and the $R^2$ and root mean square error (RMSE) values for each minimization. Note that, for SPY and IEV, the fitted value $\hat{\sigma}$ is negative; their cumulative returns have a negative dependence on $\delta^{(SPG)}$. Plots of these fits are shown (and briefly discussed) in Appendix E.

**Table 2.** The constrained least-squares minimization results from (30)

|  | SPY | IEV | JPXN |
|---|---|---|---|
| $\hat{\mu}$ | 0.32 | 0.31 | $-0.069$ |
| $\hat{\sigma}$ | $-0.090$ | $-0.23$ | 2.8 |
| $R^2$ | 0.96 | 0.89 | 0.64 |
| RMSE | 0.018 | 0.032 | 0.033 |

Consider a European put option with payoff $f_T^{(\text{put})} = \max\left(0, K - \min\left(S_T^{(1)}, S_T^{(2)}, S_T^{(3)}\right)\right)$. Using (dividend adjusted) starting prices for SPY ($432.51), IEV ($52.25) and JPXN ($76.09) for 9/14/2021, we compute the risk neutral probabilities (28) and form the risk-neutral tree (27) over a time period $[0, T]$, where $T$ is the time to maturity in days. In Fig. 4 we plot the put option price

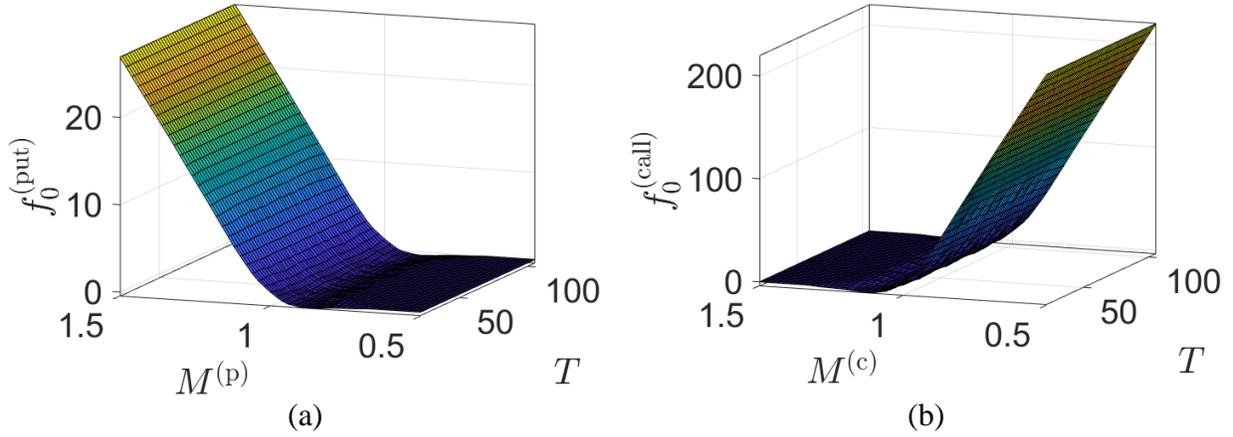

(a)                         (b)

**Figure 4.** European (a) put and (b) call option values, $f_0$, plotted against time to maturity $T$ and moneyness $M$. The moneyness values $M^{(c)}$ and $M^{(p)}$ are defined in the text.

$f_0^{(\text{put})}$ as a function of $T$ and moneyness $M^{(p)} = K/\min\left(S_0^{(1)}, S_0^{(2)}, S_0^{(3)}\right)$ over the range $T \in (0,100)$, $M^{(p)} \in (0.5,1.5)$. Repeating this for a European call with payoff $f_T^{(\text{call})} = \max\left(0, \max\left(S_T^{(1)}, S_T^{(2)}, S_T^{(3)}\right) - K\right)$, Fig. 4 also plots $f_0^{(\text{call})}$ as a function of $T$ and moneyness $M^{(c)} = K/\max\left(S_0^{(1)}, S_0^{(2)}, S_0^{(3)}\right)$, over the same range of maturity and moneyness values. The difference in price range for the put and call options represents the disparity in the computed prices of the ETFs at maturity, which in turn is based upon differences in their starting prices, drift, $\hat{\mu}$, and marginal scale parameter, $\hat{\sigma}$, values. For the specific values in Table 2, the ETFs IEV and JPXN dominate the computation of $f_0^{(\text{put})}$, while SPY dominates the computation of $f_0^{(\text{call})}$.

**Appendix A: The skew normal distribution**

**Definition**: A continuous random variable $Z^{(\lambda)}$ has a *skew normal distribution* (SND)[50] with parameter $\lambda \in R$ if its probability density function has the form $f_{Z^{(\lambda)}}(z) = 2\phi(z)\Phi(\lambda z)$, $z \in$ $R$, where $\phi$ is the standard normal density, and $\Phi$ is the standard normal cumulative distribution function.

---
[50] Azzalini (1985, 1986), Azzalini and Capitonio (2003, 2014), and Dalla Valle (2005).

Relevant properties of the SND are the following (Dalla Valle 2005).

SND 1. (a1)[51] $Z^{(\lambda)} \stackrel{d}{=} -Z^{(-\lambda)}$; (b1) $|Z^{(\lambda)}| \stackrel{d}{=} |\mathcal{N}(0,1)|$; (c1) The density of $Z^{(\lambda)}$ is unimodal; and

(d1) The moment-generating function $M_{Z^{(\lambda)}}(u) = \mathbb{E}\left[e^{uZ^{(\lambda)}}\right]$, $u \in R$, of $Z^{(\lambda)}$ is given by

$$M_{Z^{(\lambda)}}(u) = 2e^{u^2}\Phi(\delta u), \quad \delta = \lambda/\sqrt{1+\lambda^2}.$$

SND 2. Let $N_1, N_2$ be a pair of independent standard normal random variables, $N_i \stackrel{d}{=} \mathcal{N}(0,1)$, $i = 1, 2$.

(a2) If $Z = \sqrt{1-\delta^2}N_1 + \delta|N_2|$ for some $\delta \in (-1,1)$, then $Z \stackrel{d}{=} Z^{(\lambda)}$ with $\lambda = \delta/\sqrt{1-\delta^2}$.

In particular, if $\mathbb{A}^{(\delta)}$ is an Itô-Mckean SBM with parameter $\delta \in (-1,1)$, then for every $t \geq 0$, $A_t^{(\delta)} \stackrel{d}{=} \sqrt{t}Z^{(\lambda)}$, with $\lambda = \delta/\sqrt{1-\delta^2}$.

(b2) If $V = \begin{cases} N_1 & \text{when } \lambda N_1 > N_2 \\ -N_1 & \text{when } \lambda N_1 \leq N_2 \end{cases}$, then $V \stackrel{d}{=} Z^{(\lambda)}$.

SND 3. Let $(N_3, N_4)$ be a bivariate normal random variable with $N_i \stackrel{d}{=} \mathcal{N}(0,1)$, $i = 3, 4$, having correlation coefficient $\rho \in (-1,1)$. Then

(a3) the conditional distribution of $N_4$ given that $N_3 > 0$ is SND with parameter $\lambda = \rho/\sqrt{1-\rho^2}$; and

(b3) $\max(N_3, N_4) \stackrel{d}{=} Z^{(\lambda)}$ with $\lambda = \sqrt{(1-\rho)/(1+\rho)}$.

SND 4. (a4) $\mathbb{E}[Z^{(\lambda)}] = \sqrt{2/\pi}\,\delta$ and $\text{Var}(Z^{(\lambda)}) = 1 - (2/\pi)\delta^2$ with $\delta = \lambda/\sqrt{1+\lambda^2}$;

(b4) For all $k \in \mathcal{N}$, $\mathbb{E}\left[(Z^{(\lambda)})^{2k}\right] = \mathbb{E}\left[(\mathcal{N}(0,1))^{2k}\right]$ and[52]

---

[51] Here and in what follows $\stackrel{d}{=}$ stands for "equal in distribution" ("equal in probability law").
[52] Henze (1986, Proposition 1.2.3).

$$\mathbb{E}\left[\left(Z^{(\lambda)}\right)^{2k+1}\right] = \sqrt{\frac{2}{\pi}}\lambda(1+\lambda^2)^{-k-1/2}\,2^{-k}\,(2k+1)!\sum_{j=0}^{k}\frac{j!\,(2\lambda)^{2j}}{(2j+1)!\,(k-j)!}.$$

## Appendix B: Properties of the skew random walk $\mathbb{M}^{(\alpha)}$

From (2) and the fact that $\mathbb{B}_{[0,T]}^{(\alpha,n)}$ in (4) converges weakly in $\mathcal{D}[0,T]$ to the SBM $\mathbb{B}_{[0,T]}^{(\alpha)}$ as $\Delta t \downarrow 0$, we conclude that

$$\mathbb{E}\left[M_k^{(\alpha)}\right] = \mu_1^{(\alpha)}\sqrt{k}, \qquad \sqrt{\mathrm{Var}\left[M_k^{(\alpha)}\right]} = \sqrt{\left[1 - \mu_1^{(\alpha)^2}\right]k},$$
$$\mathbb{E}\left[\Delta M_k^{(\alpha)}\right] = \mu_1^{(\alpha)}\left[\sqrt{k} - \sqrt{(k-1)}\right], \quad \sqrt{\mathrm{Var}\left[\Delta M_k^{(\alpha)}\right]} = \sqrt{1 - \mu_1^{(\alpha)^2}\left[\sqrt{k} - \sqrt{(k-1)}\right]^2}, \tag{B1}$$

where $\Delta M_k^{(\alpha)} \equiv M_k^{(\alpha)} - M_{k-1}^{(\alpha)} = \pm 1$. As $\mu_1^{(0.5-\delta)} = -\mu_1^{(0.5+\delta)}$ for $\delta \in (0,0.5]$, then

$$\mathbb{E}\left[M_k^{(0.5-\delta)}\right] = -\mathbb{E}\left[M_k^{(0.5+\delta)}\right], \qquad \sqrt{\mathrm{Var}\left[M_k^{(0.5-\delta)}\right]} = \sqrt{\mathrm{Var}\left[M_k^{(0.5+\delta)}\right]},$$
$$\mathbb{E}\left[\Delta M_k^{(0.5-\delta)}\right] = -\mathbb{E}\left[\Delta M_k^{(0.5+\delta)}\right], \quad \sqrt{\mathrm{Var}\left[\Delta M_k^{(0.5-\delta)}\right]} = \sqrt{\mathrm{Var}\left[\Delta M_k^{(0.5+\delta)}\right]}. \tag{B2}$$

As we expect markets to be efficient, we are interested in the range $\alpha \in [0.4, 0.6]$. To verify (B1) for any fixed value of $\alpha$, we generated a sample of $10^6$ SRWs, each of length 6000 time-steps, and performed the expectations required in (B1) over the sample. The results for the random walk $\alpha = 0.5$ and for the SRW $\alpha = 0.6$ are shown in Fig. B1. Table B1 quantifies the root mean square error

$$\mathrm{MSE}\left(\mathbb{E}\left[M_k^{(\alpha)}\right]\right) = \sqrt{\frac{1}{6000}\sum_{k=1}^{6000}\left(\mathbb{E}\left[M_k^{(\alpha)}\right] - \mu_1^{(\alpha)}\sqrt{k}\right)^2}, \tag{B3}$$

with analogous definitions for $\mathrm{MSE}\left(\sqrt{\mathrm{Var}\left[M_k^{(\alpha)}\right]}\right)$, $\mathrm{MSE}\left(\mathbb{E}\left[\Delta M_k^{(\alpha)}\right]\right)$ and $\mathrm{MSE}\left(\sqrt{\mathrm{Var}\left[\Delta M_k^{(\alpha)}\right]}\right)$, between the ensemble averaged and the theoretical values for select values of $\alpha \in [0.5, 0.6]$. $\mathrm{MSE}\left(\mathbb{E}\left[\Delta M_k^{(\alpha)}\right]\right)$ and $\mathrm{MSE}\left(\sqrt{\mathrm{Var}\left[\Delta M_k^{(\alpha)}\right]}\right)$ grow with $\alpha$ over this range, while $\mathrm{MSE}\left(\mathbb{E}\left[M_k^{(\alpha)}\right]\right)$ and $\mathrm{MSE}\left(\sqrt{\mathrm{Var}\left[M_k^{(\alpha)}\right]}\right)$ are relatively constant.

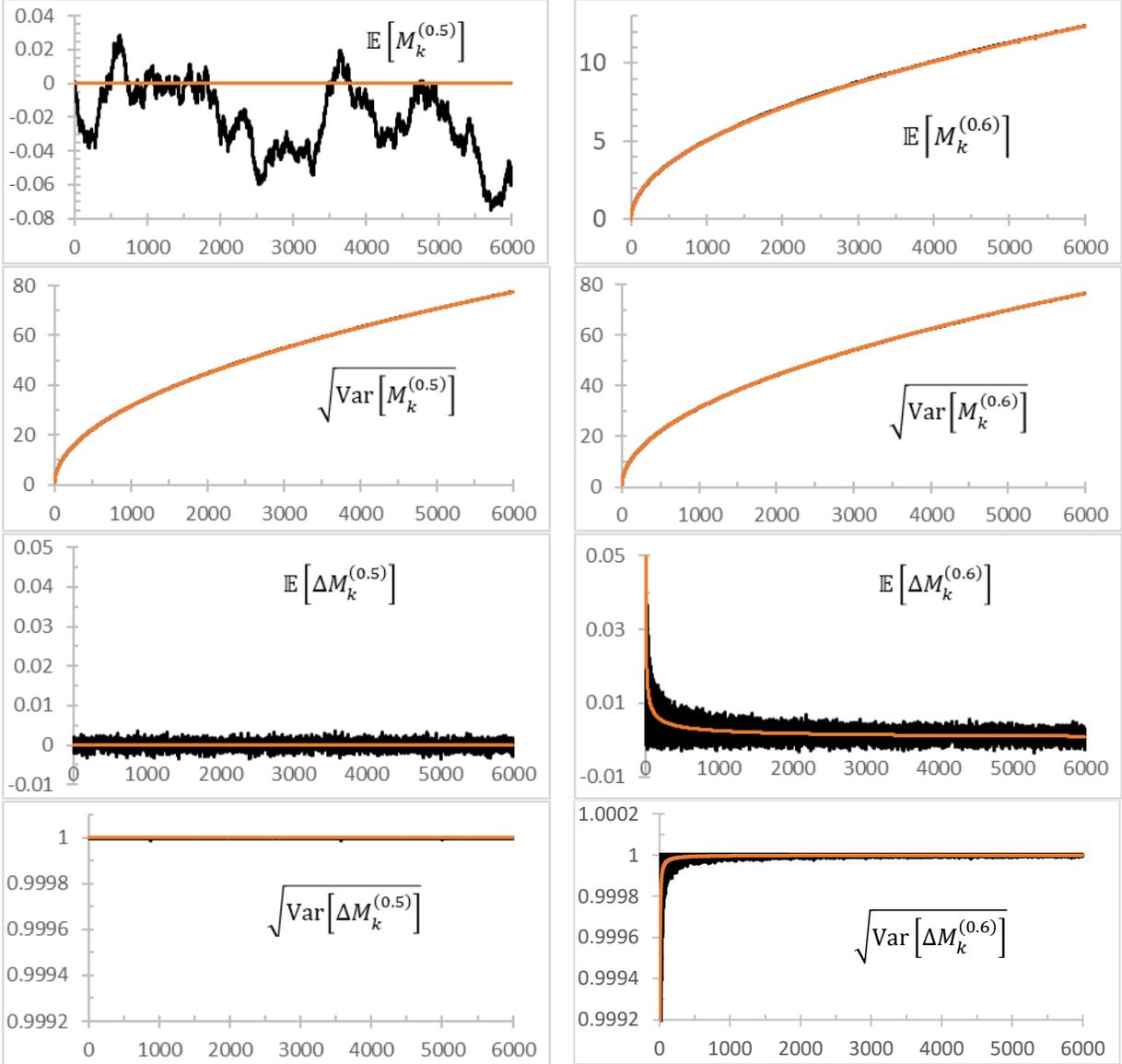

**Figure B1.** (black) Estimates of the indicated quantity from an ensemble of $10^6$ SRWs. (red) The theoretical value (B1).

**Table B1.** Root mean square error between ensemble averaged and theoretical measures

| $\alpha$ | MSE$\left(\mathbb{E}\left[M_k^{(\alpha)}\right]\right)$ | MSE$\left(\sqrt{\text{Var}\left[M_k^{(\alpha)}\right]}\right)$ | MSE$\left(\mathbb{E}\left[\Delta M_k^{(\alpha)}\right]\right)$ | MSE$\left(\sqrt{\text{Var}\left[\Delta M_k^{(\alpha)}\right]}\right)$ |
| --- | --- | --- | --- | --- |
| 0.50 | $3.0 \cdot 10^{-2}$ | $3.6 \cdot 10^{-2}$ | $1.0 \cdot 10^{-3}$ | $8.6 \cdot 10^{-7}$ |
| 0.52 | $2.6 \cdot 10^{-2}$ | $1.5 \cdot 10^{-2}$ | $1.1 \cdot 10^{-3}$ | $4.8 \cdot 10^{-6}$ |
| 0.54 | $4.7 \cdot 10^{-2}$ | $3.1 \cdot 10^{-2}$ | $1.6 \cdot 10^{-3}$ | $1.9 \cdot 10^{-5}$ |
| 0.56 | $6.8 \cdot 10^{-2}$ | $3.8 \cdot 10^{-2}$ | $2.1 \cdot 10^{-3}$ | $4.1 \cdot 10^{-5}$ |
| 0.58 | $7.8 \cdot 10^{-2}$ | $9.4 \cdot 10^{-2}$ | $2.7 \cdot 10^{-3}$ | $7.7 \cdot 10^{-5}$ |
| 0.60 | $2.5 \cdot 10^{-2}$ | $2.4 \cdot 10^{-2}$ | $3.3 \cdot 10^{-3}$ | $1.2 \cdot 10^{-4}$ |

Since the skew parameter $\alpha$ only "comes into play" at times $k\Delta t$ for which $M_k^{(\alpha)} = 0$, the question of how frequently this occurs is important for assessing the ability to determine $\alpha$ using a fitting procedure on any single trajectory. Since the trajectory of a SRW *between successive values* of $M_k^{(\alpha)} = 0$ is governed solely by a standard random walk (i.e. with up/down probabilities of ½), the distribution of the frequency of occurrence of $M_k^{(\alpha)} = 0$ is determined solely by a standard random walk, *independent of $\alpha$*. Fig. B2 confirms this by displaying the relative frequency distribution of the occurrence rate (in percent) of $M_k^{(\alpha)} = 0$ during a random walk for five choices of $\alpha$. Each distribution was determined based upon $10^6$ randomly generated $M_k^{(\alpha)}$ trajectories, each of length 6000 steps. The occurrence distribution is highly skewed, with occurrences ranging from once in 6000 steps (0.017%) to greater than 250 times (4.17%). The quartile values are $Q_1 = 0.41\%, Q_2 = 0.87\%, Q_3 = 1.48\%$.

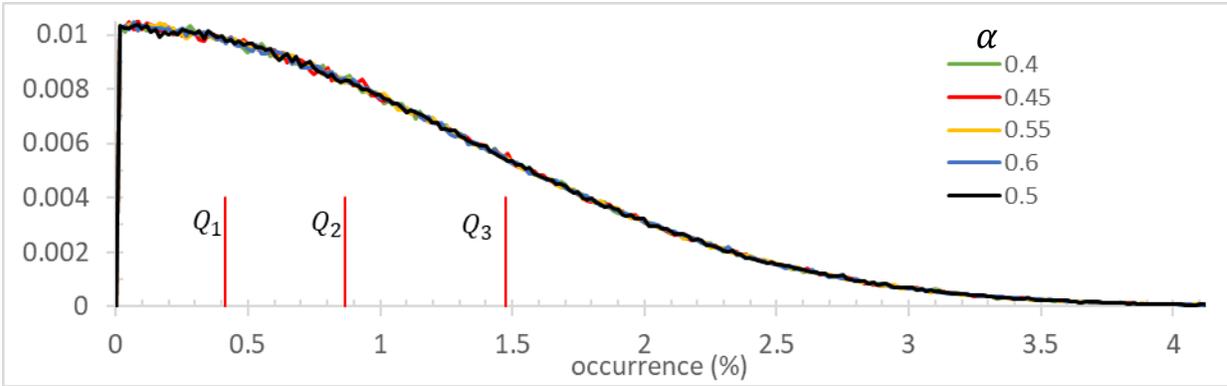

**Figure B2.** Relative frequency distribution of the occurrence rate (in percent) of $M_k^{(\alpha)} = 0$ during a SRW governed by the listed value of $\alpha$. The vertical red lines indicate the quartile values for the distribution. The distribution is truncated at 4.17%.

**Appendix C: Parameter fitting procedure**

Let $S_{k\Delta t}^{(\text{emp})}$ denote an empirical price series and $R_{k\Delta t}^{(\text{emp})}$ and $r_{k\Delta t}^{(\text{emp})}$ denote, respectively, its cumulative and daily log-returns. Fitting the price series to (5) involves solving one or both of

$$R_{k\Delta t}^{(\text{emp})} = \left(\mu - \frac{\sigma^2}{2}\right) k\Delta t + \sigma M_k^{(\alpha)} \sqrt{\Delta t} \equiv R_{k\Delta t}^{(\text{th})}(\mu, \sigma, \alpha),$$

$$r_{k\Delta t}^{(\text{emp})} = \left(\mu - \frac{\sigma^2}{2}\right) \Delta t + \sigma \Delta M_k^{(\alpha)} \sqrt{\Delta t} \equiv r_{k\Delta t}^{(\text{th})}(\mu, \sigma, \alpha),$$

(C1)

for the parameters $\mu$, $\sigma$ and $\alpha$. From (C1) and (B1) note that

$$\text{Var}\left[r_{k\Delta t}^{(\text{emp})}\right] = \sigma^2 \Delta t \, \text{Var}\left[\Delta M_k^{(\alpha)}\right] = \sigma^2 \Delta t \left(1 - \mu_1^{(\alpha)^2} \left[\sqrt{k} - \sqrt{(k-1)}\right]^2\right). \tag{C2}$$

Based upon Fig. B1 we see that the right-hand side of (C2) can be accurately approximated by

$$\text{Var}\left[r_{k\Delta t}^{(\text{emp})}\right] = \sigma^2 \Delta t, \tag{C3}$$

where the right-hand side of (C3) is independent of $k$. Thus, computing the variance of $r_{k\Delta t}^{(\text{emp})}$ over all values $k$ from a single price series provides an accurate estimation of $\sigma$. Estimating $\sigma \equiv \sigma^*$ from (C3) is the first step in our fitting procedure. We have found no method that allows useful determination of $\mu$ and $\alpha$ independently of each other, thus the second step in the parameter fitting procedure involves simultaneous estimation of $\mu$ and $\alpha$ (using the value $\sigma^*$ obtained in the first step). As numerical realizations of the SRW $M_k^{(\alpha)}$ can be computed extremely rapidly, the second step consists of generating a large ensemble of return series $R_{k\Delta t}^{(\text{th})}(\mu, \sigma^*, \alpha)$, each series based upon an independent sample pair $(\mu, \alpha) \in \mathcal{D} = [\mu_L, \mu_H] \times [\alpha_L, \alpha_H]$, and determining the parameter pair that satisfies

$$(\mu^*, \alpha^*) = \arg\min_{\mathcal{D}} \left\| R_{k\Delta t}^{(\text{emp})} - R_{k\Delta t}^{(\text{th})}(\mu, \sigma^*, \alpha) \right\|_2^2. \tag{C4}$$

Setting (or, if necessary, adjusting) the limits $\alpha_L, \alpha_H$ is relatively straightforward. For the numerical examples, we have used $\alpha_L = 0.45$, $\alpha_H = 0.65$. If the minimization produces values of $\alpha^*$ that approach either limit, the limits can be adjusted and the minimization rerun. To determine the limits $\mu_L, \mu_H$, one can consider the approximation

$$\mathbb{E}\left[r_{k\Delta t}^{(\text{emp})}\right] = \left(\mu - \frac{\sigma^{*2}}{2}\right)\Delta t + \sigma^*\mathbb{E}\left[\Delta M_k^{(\alpha)}\right]\sqrt{\Delta t} \approx \left(\mu - \frac{\sigma^{*2}}{2}\right)\Delta t, \quad (C5)$$

to obtain a rough value $\mu = \tilde{\mu}$, and then set the limits $\mu_L = \tilde{\mu} - \sigma^*/2$, $\mu_H = \tilde{\mu} + \sigma^*/2$. However, the approximation in (C5) is poor, as $\sigma^*\mathbb{E}\left[\Delta M_k^{(\alpha)}\right]\sqrt{\Delta t}$ can be comparable in value to $\left(\mu - \frac{\sigma^{*2}}{2}\right)\Delta t$. Consequently, we set the wide limits, $\mu_L = -0.5$, $\mu_H = 0.5$, which proved to be satisfactory in our minimizations.

The fitting procedure was tested on synthetic price series, each consisting of a price trajectory generated from (3) and (5), with known values of $\mu^{(\text{syn})}$, $\sigma^{(\text{syn})}$ and $\alpha^{(\text{syn})}$. Using these values, an SRW $M_k^{(\alpha)}$, k = 1, ..., 6000 was generated, from which a price trajectory $S_{k\Delta t}^{(\text{syn})}$ was computed from (5) consisting of a sequence of 6000 daily prices (5999 returns or 23.8 years of data). From $S_{k\Delta t}^{(\text{syn})}$, $R_{k\Delta t}^{(\text{syn})}$ and $r_{k\Delta t}^{(\text{syn})}$ were generated to represent the empirical data set (C1) in the fitting procedure. Table C1 presents the results of empirical fits to synthetic data sets generated with $\mu^{(\text{syn})} = 0.05$, $\sigma^{(\text{syn})} = 0.1$ and $\alpha^{(\text{syn})} = \{0.4, 0.5, 0.6\}$. For each choice of $\alpha^{(\text{syn})}$, 10 synthetic price trajectories were computed to assess accuracy by computing average and sample standard deviation of the parameter fits over the sample of 10. We see that $\sigma^*$ can be computed very accurately for any individual price series. Since the standard deviation of daily returns is larger than the value $\mu$, the term $\sigma M_k^{(\alpha)}\sqrt{\Delta t}$ in (C1) is comparable in magnitude (in our synthetic data sets, roughly 37% of the magnitude) of the drift term $(\mu - \sigma^2/2)k\Delta t$, making an estimation of $\mu$ from a *single* price trajectory much less accurate. However, averaging over as small a sample as 10 price traces leads to an average value that is within one standard deviation of the true value $\mu$. Estimation of $\alpha$ from a single price trace is slightly better than single-trace estimation of $\mu$; consequently, estimation of $\alpha$ from the small set of 10 traces yields better results than that for $\mu$. Critically, from a single price trace it is possible to infer whether $\alpha$ differs from 0.5, allowing for the possible refinement of a value for $\alpha$ by narrowing the search limits $[\alpha_\text{L}, \alpha_\text{H}]$ in the optimization procedure.

**Table C1.** Results of the application of the fitting procedure to synthetic data sets

| synthetic data set | $\alpha^{(\text{syn})} = 0.6$ | | | $\alpha^{(\text{syn})} = 0.5$ | | | $\alpha^{(\text{syn})} = 0.4$ | | |
|---|---|---|---|---|---|---|---|---|---|
| | $\sigma^*$ | $\mu^*$ | $\alpha^*$ | $\sigma^*$ | $\mu^*$ | $\alpha^*$ | $\sigma^*$ | $\mu^*$ | $\alpha^*$ |
| 1 | 0.099993 | 0.015 | 0.53 | 0.099997 | 0.000 | 0.52 | 0.099995 | 0.066 | 0.34 |
| 2 | 0.100002 | 0.061 | 0.66 | 0.099996 | 0.048 | 0.57 | 0.099994 | 0.019 | 0.34 |
| 3 | 0.100008 | 0.027 | 0.63 | 0.099990 | 0.016 | 0.50 | 0.099984 | 0.015 | 0.47 |
| 4 | 0.100008 | 0.052 | 0.65 | 0.100008 | 0.035 | 0.59 | 0.099985 | 0.083 | 0.48 |
| 5 | 0.100001 | 0.057 | 0.64 | 0.099991 | 0.010 | 0.44 | 0.100006 | 0.059 | 0.48 |
| 6 | 0.100008 | 0.043 | 0.51 | 0.099994 | 0.064 | 0.54 | 0.100000 | 0.069 | 0.36 |
| 7 | 0.100002 | 0.029 | 0.66 | 0.099998 | 0.039 | 0.56 | 0.0999471 | -0.016 | 0.47 |
| 8 | 0.099971 | 0.087 | 0.63 | 0.100003 | 0.057 | 0.54 | 0.100008 | 0.025 | 0.45 |
| 9 | 0.100007 | 0.055 | 0.52 | 0.100008 | 0.039 | 0.58 | 0.099994 | 0.008 | 0.49 |
| 10 | 0.100007 | 0.054 | 0.58 | 0.100008 | 0.019 | 0.47 | 0.099995 | 0.016 | 0.38 |
| average | 0.100001 | 0.048 | 0.60 | 0.099999 | 0.033 | 0.53 | 0.099991 | 0.035 | 0.43 |
| stdev | 0.000011 | 0.021 | 0.06 | 0.000007 | 0.021 | 0.05 | 0.000017 | 0.032 | 0.06 |

**Appendix D: The Determinants**

The determinants appearing in equation (25) are

$$\mathbb{D}_{k\Delta t} = \begin{vmatrix} \psi_k^{(1,uu,n)} - \psi_k^{(1,ud,n)} & \psi_k^{(2,uu,n)} - \psi_k^{(2,ud,n)} & \psi_k^{(3,uu,n)} - \psi_k^{(3,ud,n)} \\ \psi_k^{(1,ud,n)} - \psi_k^{(1,du,n)} & \psi_k^{(2,ud,n)} - \psi_k^{(2,du,n)} & \psi_k^{(3,ud,n)} - \psi_k^{(3,du,n)} \\ \psi_k^{(1,du,n)} - \psi_k^{(1,dd,n)} & \psi_k^{(2,du,n)} - \psi_k^{(2,dd,n)} & \psi_k^{(3,du,n)} - \psi_k^{(3,dd,n)} \end{vmatrix},$$

$$\mathbb{D}_{k\Delta t}^{(1)} = \begin{vmatrix} f_{(k+1)\Delta t}^{(n;uu)} - f_{(k+1)\Delta t}^{(n;ud)} & \psi_k^{(2,uu,n)} - \psi_k^{(2,ud,n)} & \psi_k^{(3,uu,n)} - \psi_k^{(3,ud,n)} \\ f_{(k+1)\Delta t}^{(n;ud)} - f_{(k+1)\Delta t}^{(n;du)} & \psi_k^{(2,ud,n)} - \psi_k^{(2,du,n)} & \psi_k^{(3,ud,n)} - \psi_k^{(3,du,n)} \\ f_{(k+1)\Delta t}^{(n;du)} - f_{(k+1)\Delta t}^{(n;dd)} & \psi_k^{(2,du,n)} - \psi_k^{(2,dd,n)} & \psi_k^{(3,du,n)} - \psi_k^{(3,dd,n)} \end{vmatrix},$$

$$\mathbb{D}_{k\Delta t}^{(2)} = \begin{vmatrix} \psi_k^{(1,uu,n)} - \psi_k^{(1,ud,n)} & f_{(k+1)\Delta t}^{(n;uu)} - f_{(k+1)\Delta t}^{(n;ud)} & \psi_k^{(3,uu,n)} - \psi_k^{(3,ud,n)} \\ \psi_k^{(1,ud,n)} - \psi_k^{(1,du,n)} & f_{(k+1)\Delta t}^{(n;ud)} - f_{(k+1)\Delta t}^{(n;du)} & \psi_k^{(3,ud,n)} - \psi_k^{(3,du,n)} \\ \psi_k^{(1,du,n)} - \psi_k^{(1,dd,n)} & f_{(k+1)\Delta t}^{(n;du)} - f_{(k+1)\Delta t}^{(n;dd)} & \psi_k^{(3,du,n)} - \psi_k^{(3,dd,n)} \end{vmatrix},$$

$$\mathbb{D}_{k\Delta t}^{(3)} = \begin{vmatrix} \psi_k^{(1,uu,n)} - \psi_k^{(1,ud,n)} & \psi_k^{(2,uu,n)} - \psi_k^{(2,ud,n)} & f_{(k+1)\Delta t}^{(n;uu)} - f_{(k+1)\Delta t}^{(n;ud)} \\ \psi_k^{(1,ud,n)} - \psi_k^{(1,du,n)} & \psi_k^{(2,ud,n)} - \psi_k^{(2,du,n)} & f_{(k+1)\Delta t}^{(n;ud)} - f_{(k+1)\Delta t}^{(n;du)} \\ \psi_k^{(1,du,n)} - \psi_k^{(1,dd,n)} & \psi_k^{(2,du,n)} - \psi_k^{(2,dd,n)} & f_{(k+1)\Delta t}^{(n;du)} - f_{(k+1)\Delta t}^{(n;dd)} \end{vmatrix}.$$

The terms appearing in equation (28) are the following.

$$\mathbb{q}_{(k+1)\Delta t} = \begin{vmatrix} \mathcal{B}_{k\Delta t}^{(1,1,n)} & \mathcal{B}_{k\Delta t}^{(1,2,n)} & \mathcal{B}_{k\Delta t}^{(1,3,n)} \\ \mathcal{B}_{k\Delta t}^{(2,1,n)} & \mathcal{B}_{k\Delta t}^{(2,2,n)} & \mathcal{B}_{k\Delta t}^{(2,3,n)} \\ \mathcal{B}_{k\Delta t}^{(3,1,n)} & \mathcal{B}_{k\Delta t}^{(3,2,n)} & \mathcal{B}_{k\Delta t}^{(3,3,n)} \end{vmatrix}, \quad \mathbb{q}_{(k+1)\Delta t}^{(n;uu)} = \begin{vmatrix} \mathcal{A}_{k\Delta t}^{(1,1,n)} & \mathcal{A}_{k\Delta t}^{(1,2,n)} & \mathcal{A}_{k\Delta t}^{(1,3,n)} \\ \mathcal{B}_{k\Delta t}^{(2,1,n)} & \mathcal{B}_{k\Delta t}^{(2,2,n)} & \mathcal{B}_{k\Delta t}^{(2,3,n)} \\ \mathcal{B}_{k\Delta t}^{(3,1,n)} & \mathcal{B}_{k\Delta t}^{(3,2,n)} & \mathcal{B}_{k\Delta t}^{(3,3,n)} \end{vmatrix},$$

$$\mathbb{q}_{(k+1)\Delta t}^{(n;ud)} = \begin{vmatrix} -\mathcal{A}_{k\Delta t}^{(1,1,n)} & \mathcal{B}_{k\Delta t}^{(1,2,n)} & \mathcal{B}_{k\Delta t}^{(1,3,n)} \\ \mathcal{A}_{k\Delta t}^{(1,1,n)} & \mathcal{B}_{k\Delta t}^{(2,2,n)} & \mathcal{B}_{k\Delta t}^{(2,3,n)} \\ 0 & \mathcal{B}_{k\Delta t}^{(3,2,n)} & \mathcal{B}_{k\Delta t}^{(3,3,n)} \end{vmatrix} + \begin{vmatrix} \mathcal{B}_{k\Delta t}^{(1,1,n)} & -\mathcal{A}_{k\Delta t}^{(1,2,n)} & \mathcal{B}_{k\Delta t}^{(1,3,n)} \\ \mathcal{B}_{k\Delta t}^{(2,1,n)} & \mathcal{A}_{k\Delta t}^{(1,2,n)} & \mathcal{B}_{k\Delta t}^{(2,3,n)} \\ \mathcal{B}_{k\Delta t}^{(3,1,n)} & 0 & \mathcal{B}_{k\Delta t}^{(3,3,n)} \end{vmatrix} +$$

$$+ \begin{vmatrix} \mathcal{B}_{k\Delta t}^{(1,1,n)} & \mathcal{B}_{k\Delta t}^{(1,2,n)} & -\mathcal{A}_{k\Delta t}^{(1,3,n)} \\ \mathcal{B}_{k\Delta t}^{(2,1,n)} & \mathcal{B}_{k\Delta t}^{(2,2,n)} & \mathcal{A}_{k\Delta t}^{(1,3,n)} \\ \mathcal{B}_{k\Delta t}^{(3,1,n)} & \mathcal{B}_{k\Delta t}^{(3,2,n)} & 0 \end{vmatrix},$$

$$\mathbb{q}_{(k+1)\Delta t}^{(n;du)} = \begin{vmatrix} 0 & \mathcal{B}_{k\Delta t}^{(1,2,n)} & \mathcal{B}_{k\Delta t}^{(1,3,n)} \\ -\mathcal{A}_{k\Delta t}^{(1,1,n)} & \mathcal{B}_{k\Delta t}^{(2,2,n)} & \mathcal{B}_{k\Delta t}^{(2,3,n)} \\ \mathcal{A}_{k\Delta t}^{(1,1,n)} & \mathcal{B}_{k\Delta t}^{(3,2,n)} & \mathcal{B}_{k\Delta t}^{(3,3,n)} \end{vmatrix} + \begin{vmatrix} \mathcal{B}_{k\Delta t}^{(1,1,n)} & 0 & \mathcal{B}_{k\Delta t}^{(1,3,n)} \\ \mathcal{B}_{k\Delta t}^{(2,1,n)} & -\mathcal{A}_{k\Delta t}^{(1,2,n)} & \mathcal{B}_{k\Delta t}^{(2,3,n)} \\ \mathcal{B}_{k\Delta t}^{(3,1,n)} & \mathcal{A}_{k\Delta t}^{(1,2,n)} & \mathcal{B}_{k\Delta t}^{(3,3,n)} \end{vmatrix} +$$

$$+ \begin{vmatrix} \mathcal{B}_{k\Delta t}^{(1,1,n)} & \mathcal{B}_{k\Delta t}^{(1,2,n)} & 0 \\ \mathcal{B}_{k\Delta t}^{(2,1,n)} & \mathcal{B}_{k\Delta t}^{(2,2,n)} & -\mathcal{A}_{k\Delta t}^{(1,3,n)} \\ \mathcal{B}_{k\Delta t}^{(3,1,n)} & \mathcal{B}_{k\Delta t}^{(3,2,n)} & \mathcal{A}_{k\Delta t}^{(1,3,n)} \end{vmatrix},$$

$$\mathbb{q}_{(k+1)\Delta t}^{(n;dd)} = - \begin{vmatrix} \mathcal{B}_{k\Delta t}^{(1,1,n)} & \mathcal{B}_{k\Delta t}^{(1,2,n)} & \mathcal{B}_{k\Delta t}^{(1,3,n)} \\ \mathcal{B}_{k\Delta t}^{(2,1,n)} & \mathcal{B}_{k\Delta t}^{(2,2,n)} & \mathcal{B}_{k\Delta t}^{(2,3,n)} \\ \mathcal{A}_{k\Delta t}^{(1,1,n)} & \mathcal{A}_{k\Delta t}^{(1,2,n)} & \mathcal{A}_{k\Delta t}^{(1,3,n)} \end{vmatrix},$$

and

$$\mathcal{A}_{k\Delta t}^{(1,i,n)} = e^{(r-\mu^{(i)})\Delta t} - e^{\sigma^{(i)}\left\{\sqrt{1-\delta^2}+\delta\left(\left|\zeta_k^{(2,n)}+1\right|-\left|\zeta_k^{(2,n)}\right|\right)\right\}\sqrt{\Delta t}},$$

$$\mathcal{A}_{k\Delta t}^{(2,i,n)} = e^{(r-\mu^{(i)})\Delta t} - e^{\sigma^{(i)}\left\{\sqrt{1-\delta^2}+\delta\left(\left|\zeta_k^{(2,n)}-1\right|-\left|\zeta_k^{(2,n)}\right|\right)\right\}\sqrt{\Delta t}},$$

$$\mathcal{B}_{k\Delta t}^{(1,i,n)} = \begin{pmatrix} e^{\sigma^{(i)}\left\{\sqrt{1-\delta^2}+\delta\left(\left|\zeta_k^{(2,n)}+1\right|-\left|\zeta_k^{(2,n)}\right|\right)\right\}\sqrt{\Delta t}} - \\ -e^{\sigma^{(i)}\left\{\sqrt{1-\delta^2}+\delta\left(\left|\zeta_k^{(2,n)}-1\right|-\left|\zeta_k^{(2,n)}\right|\right)\right\}\sqrt{\Delta t}} \end{pmatrix},$$

$$\mathcal{B}_{k\Delta t}^{(2,i,n)} = \begin{pmatrix} e^{\sigma^{(i)}\left\{\sqrt{1-\delta^2}+\delta\left(\left|\zeta_k^{(2,n)}-1\right|-\left|\zeta_k^{(2,n)}\right|\right)\right\}\sqrt{\Delta t}} - \\ -e^{\sigma^{(i)}\left\{-\sqrt{1-\delta^2}+\delta\left(\left|\zeta_k^{(2,n)}+1\right|-\left|\zeta_k^{(2,n)}\right|\right)\right\}\sqrt{\Delta t}} \end{pmatrix},$$

$$\mathcal{B}_{k\Delta t}^{(3,i,n)} = \begin{pmatrix} e^{\sigma^{(i)}\left\{-\sqrt{1-\delta^2}+\delta\left(\left|\zeta_k^{(2,n)}+1\right|-\left|\zeta_k^{(2,n)}\right|\right)\right\}\sqrt{\Delta t}} - \\ -e^{\sigma^{(i)}\left\{-\sqrt{1-\delta^2}+\delta\left(\left|\zeta_k^{(2,n)}-1\right|-\left|\zeta_k^{(2,n)}\right|\right)\right\}\sqrt{\Delta t}} \end{pmatrix},$$

for $i = 1,2,3$.

**Appendix E.**

Fig. E1 plots the fits to (30) for the ETFs SPY, IEV and JPXN. As noted in Table 2, the theoretical model only accounts for 64% of the variability of $R_t^{(\text{JPNX,emp})}$ over this 252-day period. In contrast, the respective $R^2$ values for IEV and SPY are 89% and 96%. The augmented Dickey-Fuller test performed on the difference series $\Delta R_t^{(i)} = R_t^{(i,\text{emp})} - R_t^{(\text{th})}$ rejects the null hypothesis of a unit root at the 5% significance level, but not at the 1% level, for both $i =$ JPXN and IEV difference series; while rejecting the null hypothesis for SPY even at the 0.1% level. We have therefore investigated fitting these $\Delta R_t^{(i)}$ series to ARFIMA(1,d,1)-GARCH(1,1)-Student $t$ models, with $d = \{0, 0.25, 0.5, 0.75, 1\}$. Analysis of these results suggests that a standard $d = 0$, ARMA(1,1)-GARCH(1,1)-Student $t$ model with 6 degrees of freedom, works well for both SPY and IEV. For JPXN, the $d = 1$ model produced the best results.

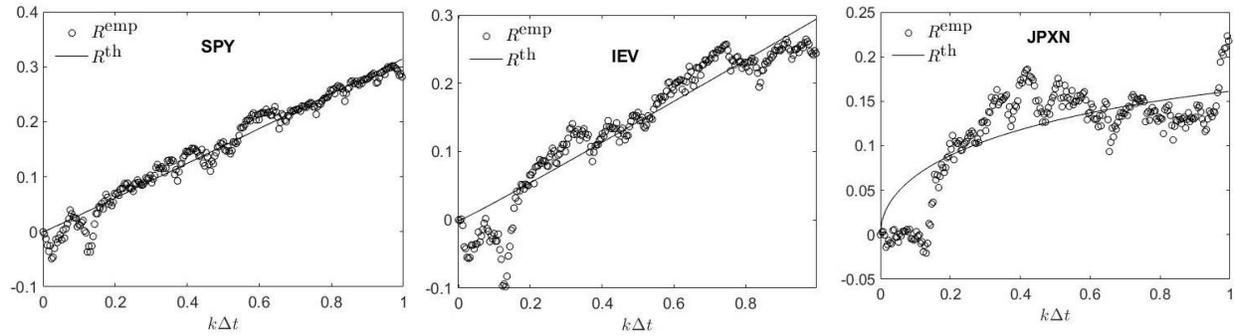

**Figure E1.** Plots of the fits to (30) for the ETFs SPY, IEV and JPXN. $R^{\text{emp}}$ denotes the empirical data, $R^{\text{th}}$ is the fitted theoretical model.